\documentclass[aps,pre,reprint,superscriptaddress,amsmath,amssymb,longbibliography,showpacs,floatfix]{revtex4-1}
\usepackage{graphicx}
\usepackage{bm}
\usepackage{color}
\usepackage{soul}
\usepackage{dcolumn}
\usepackage{multirow}
\usepackage{tikz}
\usepackage{hhline}
\usepackage{mathtools}

\begin{document}
\title{ Multidimensional entropic bound: Estimator of entropy production for Langevin dynamics with an arbitrary
time-dependent protocol }

\author{Sangyun Lee}
\affiliation{School of Physics, Korea Institute for Advanced Study, Seoul, 02455, Korea}

\author{Dong-Kyum Kim}
\affiliation{Data Science Group, Institute for Basic Science, Daejeon, 34126, Korea}

\author{Jong-Min Park}
\affiliation{School of Physics, Korea Institute for Advanced Study, Seoul, 02455, Korea}

\author{Won Kyu Kim}
\affiliation{School of Computational Sciences, Korea Institute for Advanced Study, Seoul, 02455, Korea}

\author{Hyunggyu Park}
\affiliation{School of Physics, Korea Institute for Advanced Study, Seoul, 02455, Korea}
\affiliation{Quantum Universe Center, Korea Institute for Advanced Study, Seoul 02455, Korea}

\author{Jae Sung Lee}
\email[]{jslee@kias.re.kr}
\affiliation{School of Physics, Korea Institute for Advanced Study, Seoul, 02455, Korea}

\date{\today}

\begin{abstract}
Entropy production (EP) is a key quantity in thermodynamics, and yet measuring EP has remained a challenging task. Here we introduce an EP estimator, called multidimensional entropic bound (MEB), utilizing an ensemble of trajectories. The MEB can accurately estimate the EP of overdamped Langevin systems with an arbitrary time-dependent protocol. Moreover, it provides a unified platform to accurately estimate the EP of underdamped Langevin systems under certain conditions.
In addition, the MEB is computationally efficient because optimization is unnecessary. We apply our developed estimator to three physical systems driven by time-dependent protocols pertaining to experiments using optical tweezers: a dragged Brownian particle, a pulling process of a harmonic chain, and an unfolding process of an RNA hairpin. Numerical simulations confirm the validity and efficiency of our method.
\end{abstract}

\maketitle

\section{introduction}
Entropy production (EP), referring to the quantification of the irreversibility of a thermodynamic process, is one of the most fundamental thermodynamic quantities. The EP was originally identified in the Clausius form in equilibrium thermodynamics. More recently, crucial progress in the field of thermodynamics has been the extension of the EP to general nonequilibrium phenomena at the level of a single stochastic trajectory.
This extension triggered a renaissance of thermodynamics, namely the establishment of stochastic thermodynamics. Based on the novel EP formulation, EP theories have been developed and extensively studied over the last two decades. An early one is the fluctuation theorem~\cite{Jarzynski1997nonequilibrium, Crooks1999entropy, Seifert2019stochastic, Seifert2005IFT, Collin2005verify}, which can be understood as a generalization of the thermodynamic second law. Later developments include a group of thermodynamic trade-off relations such as the thermodynamic uncertainty relation (TUR)~\cite{barato2015thermodynamic, Gingrich2016Dissipation, Horowitz2017proof, Dechant2018current, Hasegawa2019uncertainty, Dechant2019multidimensional, Hasegawa2019, Koyuk2020}, the power-efficiency trade-off relation~\cite{Benenti2011, Brandner2013, Shiraish2016, Dechant2018entropic, JSLee2020}, and the speed limit~\cite{Shiraishi2018, Nicholson2020, Hasegawa2020PRE, Esposito2020, Ito2021PRL, JSLee2022Landauer}.

Subsequently, experimentally feasible methods for measuring the EP have been actively suggested and discussed~\cite{Roldan2010Estimating,roldan2012entropy, battle2016broken,Avinery2019Universal,martiniani2019quantifying,Li2019quantifying, manikandan2019inferring,martinez2019inferring, kim2020neep,otsubo2020estimating,otsubo2020time,skinner2021wtd, kim2021estimating,van2022thermodynamic,Jarzynski2020}. In fact, measuring EP is not a trivial task. It is almost impossible to measure the EP by using its definition, the logarithmic ratio of forward and time-reversal path probabilities~\cite{seifert2012stochastic}, since all path probabilities cannot be measured, especially for a continuous system. Instead, there exist two \emph{direct} EP measurement methods using the ``equality'' for the total EP, $\Delta S^{\rm tot}$. The first method uses the equality $\Delta S^{\rm tot} = \Delta S^{\rm sys} + Q/T$, where $\Delta S^{\rm sys}$ is the Shannon entropy change of a system and $Q$ is dissipated heat into a reservoir at temperature $T$~\cite{sekimoto2010stochastic,seifert2012stochastic}. In experiments, it is difficult to measure the amount of heat flow accurately with a calorimeter. One may calculate $Q$ from trajectory data instead, but this requires full knowledge of the external and internal forces acting on the system~\cite{sekimoto2010stochastic}. Therefore, this method is not practically useful for complicated cases such as a biological system.
The second direct method uses the equality for the average EP in terms of the probability density function~(PDF) and the irreversible probability current as presented in Eq.~\eqref{eq:total_EPrate} of Ref.~\cite{Li2019quantifying}. The PDF and the irreversible probability current can be estimated solely from system trajectories without knowledge of applied forces in the overdamped Langevin dynamics. Nevertheless, obtaining them precisely for a high-dimensional system is infeasible in practice, which is called the ``curse of dimensionality''.

To overcome these shortcomings of the direct methods, several \emph{indirect} methods using a thermodynamic ``inequality'' have been suggested. Here, the EP can be estimated from an ensemble of system trajectories, and the curse of dimensionality can be mitigated by measuring several observable currents only. The indirect methods are based on an inequality in the general form of $\Delta S^{\rm tot} \geq B(\Theta)$, where the EP bound $B(\Theta)$ is determined by an observable current $\Theta$. In a certain condition, one can find an optimal observable current $\Theta^*$, which yields $B(\Theta^*) = \Delta S^	{\rm tot}$. Then, the EP can be accurately estimated by measuring $\Theta^*$.

Regarding the above indirect methods, there exist two representative inequalities. The first inequality is in TUR form~\cite{barato2015thermodynamic, Gingrich2016Dissipation, Horowitz2017proof, Dechant2018current, Hasegawa2019uncertainty}, 
where the EP is bounded by the relative fluctuation of a certain observable current. To access a tighter bound of this TUR, multidimensional TUR~\cite{Dechant2019multidimensional} and Monte Carlo methods~\cite{Li2019quantifying} have been developed. However, TURs depend on the nature of the system dynamics; e.g, the TUR must be modified when a time-dependent protocol is involved~\cite{Koyuk2020} or when a system follows underdamped Langevin dynamics~\cite{Lee2019Thermodynamic,Hasegawa2019,lee2021universal}. Thus, EP estimation based on TURs is not universal.  Moreover, if we use a TUR for an underdamped system, EP estimation is not possible from only system trajectories, but rather needs full knowledge and full controllability of the applied forces~\cite{Lee2019Thermodynamic,Hasegawa2019,lee2021universal}.
Thus, no proper method via TUR exists for estimating the EP solely from system trajectories for underdamped dynamics.

The second inequality for the indirect method is the Donsker--Varadhan inequality~\cite{donsker1983varadhan}. Recently, a machine learning technique named NEEP (neural estimator for entropy production)~\cite{kim2020neep, otsubo2020estimating, kim2021estimating} utilized this inequality as an optimized function for a given neural network. Though this technique yields a reliable result in overdamped Langevin systems, a high computational cost is required for a process with a time-dependent protocol since the parameters of the neural network should be reoptimized every single time. Otherwise, this machine learning technique has also been applied to underdamped Langevin dynamics; however, it has difficulty in estimating the EP accurately for large inertia~\cite{kim2021estimating}.

In this study, we propose a unified and computationally efficient method to estimate the EP by using the entropic bound (EB) inequality introduced by Dechant and Sasa~\cite{Dechant2018entropic}. Inspired by the multidimensional TUR~\cite{Dechant2019multidimensional}, we use multiple observable currents to obtain the optimal EB for the EP. Thus, we call this the ``multidimensional entropic bound'' (MEB). The MEB is universal in the sense that it provides a unified platform to estimate the EP for both overdamped and underdamped systems regardless of the time dependence of the driving protocol. The MEB can estimate the EP from system-trajectory information when an irreversible force is absent, a common experimental setup. When an irreversible force is involved, additional information about the force is required to estimate the EP. For an underdamped system, supplementary information on the velocity relaxation time, which can be determined experimentally, is necessary to estimate the EP.

This paper is organized as follows.
In Sec.~\ref{sec:MEB}, we derive the formulae for the MEB and describe the EP estimation process using it. In Sec.~\ref{sec:comparison}, we explain the relation between the MEB and various TUR bounds. In Sec.~\ref{sec:verification}, we apply the MEB to three systems with time-dependent driving forces that can be realized in experiments using optical tweezers. We conclude the paper in Sec.~\ref{sec:disscussion}.

\section{Multidimensional Entropic Bound}
\label{sec:MEB}

The EB is the inequality between the EP  and an observable current~\cite{Dechant2018entropic}. As this bound holds for both overdamped and underdamped Langevin systems with an arbitrary time-dependent protocol, it can be a good starting point to obtain a unified and efficient EP estimator applicable to both overdamped and underdamped Langevin dynamics. In this section, we introduce the multidimensional entropic bound (MEB) estimator by
incorporating multiple observable currents systematically into the EB estimator.

\subsection{Derivation of the integral and the rate EB} \label{sec:EBderivation}

Here, we consider an  $M$-dimensional  Langevin system with a state vector ${\bm q} (t)= (q_1, \cdots, q_M)^\textsf{T}$, where $\textsf T$ denotes the transpose of a matrix,  described by the following equation of motion:
\begin{align}
    \dot{\bm q} (t) = \bm A (\bm q(t), t ) + \sqrt{2\mathsf B (\bm q(t), t )} \bullet \bm \xi(t),
\label{eq:genLangevin}
\end{align}
where $\bm A= (A_1, \cdots, A_M)^{\textsf T}$ is a time-dependent drift force, $\mathsf B$ is a positive-definite symmetric $M \times M$ diffusion matrix, and $\bm \xi = (\xi_1, \cdots, \xi_M)^{\textsf T}$  is a Gaussian white noise satisfying $\langle \xi_i(t) \xi_j(t')\rangle = \delta_{ij}\delta(t-t')$ for $i, j\in \{1,\cdots,M\}$.  The symbol $\bullet$ in Eq.~\eqref{eq:genLangevin} represents the It\^{o} product. From now on, we sometimes drop the arguments of functions for simplicity.

A component of $\bm q$ can be an odd-parity variable such as a velocity under time-reversal operation. The time reversal of a state $\bm q$ is  denoted by $\tilde{\bm q} = (\tilde{q}_1, \cdots, \tilde{q}_M)^\textsf T$ with  $\tilde{q}_i = \epsilon_i q_i$, where $\epsilon_i=1$ for an even-parity variable and $\epsilon_i=-1$ otherwise.
The drift force can be divided into reversible and irreversible parts as $\bm A (\bm q, t ) = \bm A^{\rm rev}(\bm q, t) + \bm A^{\rm irr}(\bm q, t) $ with~\cite{Risken}
\begin{align}
	\bm A^{\rm rev}(\bm q, t) &\equiv  \frac{1}{2} \left[ \bm A(\bm q, t) - \bm \epsilon \odot \bm  A^\dagger(\bm \epsilon\odot \bm q,t) \right],\nonumber \\
	\bm A^{\rm irr}(\bm q, t) &\equiv \frac{1}{2} \left[ \bm A(\bm q, t) + \bm \epsilon \odot \bm  A^\dagger(\bm \epsilon \odot \bm q,t) \right], \label{eq:rev_irr}
\end{align}
where $\bm \epsilon = (\epsilon_1, \cdots, \epsilon_M)^\textsf T$, $\odot$ denotes the element-wise product, i.e.,~$\bm a \odot \bm b  = ( \cdots ,a_i b_i, \cdots )^\textsf T $, and $\dagger$ is an operation changing the sign of the odd-parity parameters.

The Fokker--Planck equation associated with Eq.~\eqref{eq:genLangevin} is
\begin{align}
    \partial_t P (\bm q, t) = -\bm \nabla [ \bm J^{\rm rev}(\bm q, t) + \bm J^{\rm irr}(\bm q, t) ]~,
 \end{align}
with the PDF $P(\bm q, t)$. The reversible current $\bm J^{\rm rev}(\bm q, t)$ and the irreversible current $\bm J^{\rm irr}(\bm q, t)$ are defined as
\begin{align}
    J_i^{\rm rev}(\bm q, t) &\equiv A_i^{\rm rev}(\bm q, t)P(\bm q,t ), \label{eq:Jrev} \\
    J_i^{\rm irr}(\bm q, t ) &\equiv A_i^{\rm irr} (\bm q, t) P(\bm q, t)- \sum_j \partial_{q_j} \left[ \mathsf B_{ij}(\bm q, t)  P(\bm q, t ) \right] , \label{eq:Jirr}
\end{align}
with $\mathsf B(\bm \epsilon \odot \bm q, t) =\mathsf B(\bm q, t)$.
Note that for an overdamped Langevin system with even-parity variables only, $\bm A^{\rm rev}(\bm q, t)$ vanishes, and thus $\bm J^{\rm rev}(\bm q, t)=0$ and the total current coincides with $\bm J^{\rm irr}(\bm q, t )$. As dissipation originates from the irreversible current, the EP is determined only by $\bm J^{\rm irr}(\bm q, t )$. Therefore, the total EP rate $\sigma^{\rm tot}$ is given by~\cite{Spinney2012PRE,Kwon2016JKPS,Dechant2018entropic}
\begin{align}
	\sigma^{\rm tot} (t) \equiv  \int d\bm q \frac{{\bm J}^{\rm irr} (\bm q, t)^\textsf T \mathsf B (\bm q, t)^{-1}  {\bm J}^{\rm irr} (\bm q, t) }{P(\bm q, t)}.
	\label{eq:total_EPrate}
\end{align}
Hereafter, we use the $k_B=1$ unit.
Note that for underdamped Langevin systems, the matrix is not directly invertible since $\mathsf B_{ij}=0$ when the component index $i$ or $j$ denotes a positional variable. For such an index $i$, we first set $\mathsf B_{ii}=b$ ($b>0$) and $\mathsf B_{ij}=0$ ($i\neq j$), then take the inverse of $\mathsf B$ and calculate ${\bm J}^{\rm irr} (\bm q, t)^\textsf T \mathsf B (\bm q, t)^{-1}  {\bm J}^{\rm irr} (\bm q, t)$ in Eq.~\eqref{eq:total_EPrate}, and finally take the $b \rightarrow 0$ limit. Since  $ J_i^{\rm irr} (\bm q, t) \propto b$, this limit leads to ${ J}_i^{\rm irr} (\bm q, t)^2 \mathsf B_{ii} (\bm q, t)^{-1} \sim b \rightarrow 0 $. For underdamped systems, this procedure amounts to writing $\textsf B$ and ${\bm J}^{\rm irr}$ in terms of velocity-variable components only.

In this study, we consider the following form of an averaged observable current generated by the irreversible current during time $\tau$:
\begin{align}
	\langle \Theta (\tau) \rangle = \int^\tau_0 dt\int d\bm q ~\bm \Lambda (\bm q,t )^\textsf T   \bm J^{\rm irr}(\bm q,t), \label{eq:observable}
\end{align}
where $\bm \Lambda(\bm q, t) = (\Lambda_1, \cdots, \Lambda_M)^\textsf T$ is a weight vector of the irreversible current for a given observable. Then, the averaged current rate at time $t$ is given as
\begin{align}
	\langle \dot{\Theta} (t) \rangle = \int d\bm q ~\bm \Lambda (\bm q,t)^\textsf T   \bm J^{\rm irr}(\bm q,t)~. \label{eq:observable_rate}
\end{align}

The EB  in an integral form can be derived from Eq.~\eqref{eq:observable} as follows:
\begin{align}
	&\langle \Theta (\tau) \rangle \nonumber \\
	&=  \int^\tau_0 dt\int d\bm q P(\bm q, t)^{\frac{1}{2}}\bm \Lambda (\bm q,t )^\textsf T \mathsf B(\bm q, t)^{\frac{1}{2}} \frac{\mathsf B(\bm q, t)^{-\frac{1}{2}}   \bm J^{\rm irr}(\bm q,t)}{P(\bm q, t)^{\frac{1}{2}}} \nonumber \\	
	&\leq   \sqrt{\int^\tau_0 dt \left\langle \bm \Lambda^\textsf T  \mathsf B  \bm \Lambda   \right\rangle_{\bm q}} \sqrt{\Delta S^{\rm tot} (\tau) },
	\label{eq:EB_integral}
\end{align}
where $\langle \cdots \rangle_{\bm q} = \int d \bm q \cdots P(\bm q, t) $ and the total EP $\Delta S^{\rm tot} (\tau) = \int_0^\tau dt~ \sigma^{\rm tot} (t)$. The Cauchy--Schwartz inequality is used for the last inequality of Eq.~\eqref{eq:EB_integral}. Hence, the total EP is bounded in an integral form as
\begin{align}
	\Delta S^{\rm tot} (\tau) \geq \frac{\langle \Theta (\tau) \rangle^2}{ \int^\tau_0 ds \left\langle \bm \Lambda^\textsf T  \mathsf B  \bm \Lambda   \right\rangle_{\bm q}} \equiv &\Delta S^{\rm EB} (\Theta,\tau). \nonumber  \\
	&~~~~~\textrm{(integral EB)}
	\label{eq:EB_integral1}
\end{align}
Similarly, the EB in a rate form can also be obtained from Eq.~\eqref{eq:observable_rate} as
\begin{align}
	 \sigma^{\rm tot} (t) \geq \frac{\langle \dot{\Theta} (t) \rangle^2}{ \left\langle \bm \Lambda^\textsf T  \mathsf B  \bm \Lambda   \right\rangle_{\bm q}} \equiv \sigma^{\rm EB} (\Theta,t) . ~~\textrm{(rate EB)}
	\label{eq:EB_rate}
\end{align}

The equality of the integral EB is satisfied when the weight vector has the following form:
\begin{align}
	\bm \Lambda^{\rm e} (\bm q, t) = c\frac{\mathsf B(\bm q, t)^{-1}  \bm J^{\rm irr}(\bm q,t)}{P(\bm q, t)}~~\textrm{(for the integral EB)}, \label{eq:weight_integral}
\end{align}
where $c$ is an arbitrary constant that is independent of $\bm q$ and $t$. This can be easily checked by inserting Eq.~\eqref{eq:weight_integral} into Eq.~\eqref{eq:EB_integral1}.  The weight vector in this case corresponds to the observable current
proportional to the total EP, i.e.,~$\langle \Theta (\tau)\rangle= c \Delta S^{\rm tot} (\tau)$.  Similarly, we find the equality condition for the rate EB as
\begin{align}
	\bm \Lambda^{\rm e} (\bm q, t) = c(t) \frac{\mathsf B(\bm q, t)^{-1}   \bm J^{\rm irr}(\bm q,t)}{P(\bm q, t)}~~\textrm{(for the rate EB)}, \label{eq:weight_rate}
\end{align}
where $c(t)$ is an arbitrary time-dependent function that is independent of $\bm q$.
This weight vector corresponds to the observable current rate as $\langle \dot{\Theta} (t)\rangle= c(t) \sigma^{\rm tot} (t)$.
Note that $c$ and $c(t)$ can be arbitrary, and thus we may choose $c$ and $c(t)$ freely in order to simplify the measurement of an observable current. A relevant example is presented in Sec.~\ref{sec:example1}.

\subsection{Derivation of the integral and the rate MEB} \label{sec:MEBderivation}

With the knowledge of the functional form of $\bm \Lambda^{\rm e} (\bm q, t)$, one may obtain the tight EP bound.
However, except for very simple examples, it is impossible to identify $\bm \Lambda^{\rm e} (\bm q, t)$ without knowing
all driving  and interaction forces. Instead, we measure multiple observable currents to access a tighter bound, thereby
systematically approaching the total EP. Our MEB method is analogous to the multidimensional TUR~\cite{Dechant2019multidimensional} but
is more general in the sense that it can be applicable to wider classes of Langevin dynamics.

In this method, a linear combination of multiple  weight vectors is adopted to approximate $\bm \Lambda^{\rm e} (\bm q, t)$.
The linear combination of $\ell$ weight vectors $\{\Lambda_{i,1}, \cdots, \Lambda_{i,\ell} \}$ for the $i$-th component is written  as
\begin{align}
	\Lambda_i^{(\ell)} (\bm q, t) = \sum_{\alpha=1}^\ell k_{i,\alpha} \Lambda_{i,\alpha} (\bm q, t), \label{eq:linearComb}
\end{align}
where $k_{i,\alpha}$ is the coefficient for $\Lambda_{i,\alpha} (\bm q, t)$ and is independent of $\bm q$ and $t$.
From now on, we will consider the case $\mathsf B_{ij} = B_i \delta_{ij}$ for simplicity. Even when the diffusion matrix has off-diagonal elements, we can always diagonalize the diffusion matrix by using a proper transformation of the coordinate if the full information of $\mathsf B_{ij}$ is given. Thus, we can still set $\mathsf B_{ij} = B_i \delta_{ij}$ on the transformed coordinate. In cases where it is difficult to obtain the information of $\mathsf B_{ij}$, and thus not possible to find the proper transformation, we cannot use the following MEB in component-wise form. However, even in such cases, we can still derive the MEB in ``component-combined'' form as we show in Appendix~\ref{appendix:non-component-wse MEB}.
The observable in Eq.~\eqref{eq:observable} can be divided into the sum of  its components as $\langle \Theta (\tau) \rangle = \sum_{i=1}^M \langle \Theta_i (\tau)\rangle $, where $\langle \Theta_i (\tau) \rangle = \int^\tau_0 dt\int d\bm q ~ \Lambda_i (\bm q,t )    J_i^{\rm irr}(\bm q,t)$.
With the $i$-th component current $\langle \Theta_i (\tau) \rangle$, we derive the component-wise EB as
\begin{align}
	\Delta S_i (\tau) \geq
	\frac{\langle \Theta_i (\tau) \rangle^2}{ \int^\tau_0 dt \left\langle \Lambda_i   B_i  \Lambda_i   \right\rangle_{\bm q}}
	. ~~\textrm{($i$-th integral EB)},
	\label{eq:EB_integral_component}
\end{align}
where $\Delta S_i (\tau) = \int_0^\tau dt \sigma_i (t)$ with the $i$-th component EP rate $\sigma_i(t)=\int d\bm q~   B_i(\bm q, t)^{-1} J_i^{\rm irr} (\bm q,t)^2/ P(\bm q, t)$. Thus, $\Delta S^{\rm tot}(\tau) = \sum_i \Delta S_i (\tau) $ and $\sigma^{\rm tot} (t) = \sum_i \sigma_i (t) $.
By substituting Eq.~\eqref{eq:linearComb} into Eq.~\eqref{eq:EB_integral_component}, we have
\begin{align}
    \Delta S_i (\tau) \geq \frac{ \left\{\bm k_i ^{\textsf T} {\langle \bm \Theta }_i^{(\ell)} (\tau) \rangle \right\}^2 }{\bm k_i^{\textsf T}  \mathsf L_i^{(\ell)} (\tau)  \bm k_i } \equiv \Delta \hat{S}_i^{(\ell)} (\bm k_i),
    \label{eq:combEB}
\end{align}
where $\bm k_i=(k_{i,1}, \cdots, k_{i,\ell})^{\textsf T}$, and the vector $\langle\bm \Theta_i^{(\ell)}(\tau)\rangle = (\langle\Theta_{i,1} (\tau)\rangle, \cdots, \langle\Theta_{i,\ell} (\tau)\rangle)^{\textsf T}$ and the $\ell \times \ell$ matrix $\mathsf L_i (\tau)$   are defined as
\begin{align}
  \langle \Theta_{i,\alpha} (\tau)\rangle \equiv& \int^\tau_0 dt \int d\bm q ~  \Lambda_{i, \alpha} (\bm q, t)   J_i^{\rm irr }(\bm q,t )~~ \textrm{and}  \label{eq:Theta_ialpha} \\
  \left( \mathsf L_i^{(\ell)} (\tau )\right)_{\alpha,\beta} \equiv& \int^\tau_0 dt \left\langle \Lambda_{i, \alpha} (\bm q, t)   B_i (\bm q, t)   \Lambda_{i,\beta} (\bm q, t) \right\rangle_{\bm q}, \label{eq:Lmat}
\end{align}
respectively.
Note that  $\mathsf L_i^{(\ell)} (t)$ is a positive definite matrix since $\bm z^{\textsf T}  \mathsf L_i^{(\ell)} (\tau ) \bm  z =\int^\tau_0 dt  \int d\bm q \langle \left(\sum_\alpha \sqrt{ B_i} \Lambda_{i,\alpha} z_\alpha \right)^2 \rangle_{\bm q}  >0$ for an arbitrary $\bm z $.

The bound $\Delta \hat{S}_i^{(\ell)} (\bm k_i)$ in Eq.~\eqref{eq:combEB} is a function of $\bm k_i$; thus, the tightest bound can be written as  $\Delta \hat{S}_i^{(\ell)} (\bm k_i^*)$, where $\bm k_i^*$ is the optimal vector maximizing the bound. The optimal vector is obtained by solving the following equation:
\begin{align}
	&\partial_{k_{i,\alpha} }  \Delta \hat S_{i}^{(\ell)} (\bm k_i) = 0 \nonumber \\
	& = \frac{2 \bm k_i^{\textsf T} \langle\bm \Theta_i^{(\ell)}\rangle \left\{  \bm k_i^{\textsf T}  {\mathsf L}_i^{(\ell)}   \bm k_i \cdot \langle\Theta_{i,\alpha} \rangle - \bm k_i^{\textsf T} \langle \bm \Theta_i^{(\ell)}\rangle \cdot  ({\mathsf L}_i ^{(\ell)}  \bm k_i)_\alpha \right\} }{(\bm k_i^{\textsf T}  \mathsf L_i^{(\ell)}  \bm k_i)^2} . \label{eq:optimal}
\end{align}
We can easily check that the numerator vanishes with the choice of
$\bm k_i^* = \mathsf (L_i^{(\ell)})^{-1} \cdot \langle\bm \Theta^{(\ell)}_i(\tau)\rangle$.
By plugging $\bm k_i^*$ into Eq.~\eqref{eq:combEB}, we find the component-wise MEB as
\begin{align}
    \Delta S_i (\tau) \geq \langle \bm \Theta_i^{(\ell)} (\tau) \rangle^{\textsf T}   \mathsf (L_i^{(\ell)} (\tau))^{-1}   \langle \bm \Theta_i^{(\ell)} (\tau) \rangle \equiv \Delta  S_i^{{\rm MEB} (\ell)} (\tau)
    \label{eq:partialEB}
.\end{align}
By summing over all components, we finally obtain our main result, namely MEB in integral form, as follows:
\begin{align}
	\Delta S^{\rm tot} (\tau) \geq &\sum_{i=1}^M \Delta  S_i^{{\rm MEB} (\ell)} (\tau) \nonumber \\
	&\equiv \Delta  S^{{\rm MEB}(\ell)} (\tau). ~~~\textrm{(integral MEB)}
	\label{eq:integralMEB}
\end{align}

We can also derive the MEB in rate form. The derivation of the rate MEB is essentially the same as that of the integral MEB. It starts from the component-wise rate EB as
\begin{align}
	\sigma_i (t) \geq \frac{\langle \dot{\Theta}_i (t) \rangle^2}{ \left\langle   \Lambda_i^\textsf T  \mathsf B_i   \Lambda_i   \right\rangle_{\bm q}}~. ~~~~~\textrm{($i$-th rate EB)} \label{eq:ith_rateEB}
\end{align}
In this case, it is usually sufficient to choose a time-independent basis as
\begin{align}
	\Lambda_i^{(\ell)} (\bm q, t) = \sum_{\alpha=1}^\ell k_{i,\alpha} (t) \Lambda_{i,\alpha} (\bm q), \label{eq:linearComb_rate}
\end{align}
where the time-dependence is encoded in the coefficients instead of in $\Lambda_{i,\alpha} (\bm q)$, as the equality condition in Eq.~\eqref{eq:weight_rate}
also allows a time-dependent overall multiplicative coefficient.
Following the same derivation procedure as in Eqs.~\eqref{eq:combEB}$-$\eqref{eq:integralMEB}, we finally obtain
\begin{align}
	\sigma^{\rm tot} (t) \geq &\sum_{i=1}^M \langle  \dot{\bm \Theta}_i^{(\ell)} (t) \rangle^{\textsf T}   \left(\dot{\mathsf L}_i^{(\ell)} (t) \right)^{-1}   \langle \dot{\bm \Theta}_i^{(\ell)} (t) \rangle \nonumber\\
	=&\sum_{i=1}^M \sigma_i^{{\rm MEB}(\ell)} (t)\nonumber \\
	 \equiv& \sigma^{{\rm MEB}(\ell) }(t), ~~~~~~~~~~~~~~\textrm{(rate MEB)}
	\label{eq:rateMEB}
\end{align}
where  the vector $\langle\dot{\bm \Theta}_i^{(\ell)}(t)\rangle = (\langle \dot{\Theta}_{i,1} (t)\rangle, \cdots, \langle\dot{\Theta}_{i,\ell} (t)\rangle)^{\textsf T}$ and the $\ell \times \ell$ matrix $\dot{\mathsf L}_i^{(\ell)} (\tau)$   are defined as
\begin{align}
	\langle \dot{\Theta}_{i,\alpha} (t)\rangle \equiv&  \int d\bm q ~  \Lambda_{i, \alpha} (\bm q)   J_i^{\rm irr }(\bm q,t )~~ \textrm{and}  \label{eq:Theta_rate} \\
	\left( \dot{\mathsf L}_i^{(\ell)} (t) \right)_{\alpha,\beta} \equiv&  \left\langle \Lambda_{i, \alpha} (\bm q)   B_i (\bm q, t)   \Lambda_{i,\beta} (\bm q) \right \rangle_{\bm q}. \label{eq:Lmat_rate}
\end{align}
The total EP during a finite time $\tau$ can be evaluated by integrating $\sigma^{{\rm MEB}(\ell)} (t)$ over time.

The weight vectors for the rate MEB are not time-dependent, so we need a lower number of weight vectors to approximate $\Lambda_i^{\rm e} (\bm q, t)$ compared to the integral MEB where time-dependent weight vectors are necessary.
Practically, too many weight vectors can overfit all the fluctuations originating from a finite number of trajectories, sometimes giving rise to an undesirable result. Thus, the rate MEB is usually preferable in estimating the EP for a system driven by a time-dependent protocol.


The MEBs in Eqs.~\eqref{eq:integralMEB} and \eqref{eq:rateMEB} are the maximum bounds for a given finite number of observables. If we add one more observable to the existing $\ell$ observables, the MEB becomes tighter, i.e.,~
\begin{align}
	\Delta S_i^{{\rm MEB}(\ell+1)}(\tau) - \Delta  S_i^{{\rm MEB}(\ell)} (\tau) &\geq 0,  \nonumber\\
	\sigma_i^{{\rm MEB}(\ell+1)}(t) - \sigma_i^{{\rm MEB}(\ell)} (t) &\geq 0.
	\label{eq:increasing_l}
\end{align}
The proof of Eq.~\eqref{eq:increasing_l} is basically the same as that presented in Ref.~\cite{dechant2021improving}. To be self-contained, we include the proof in Appendix~\ref{appendix:MEBandaddingbasis}.
As we increase $\ell$, the MEB estimator also increases and eventually saturates to the maximum value, i.e.,~$\Delta S_i (\tau)$ or  $\sigma_i (t)$. It can often saturate even at finite $\ell=\ell^{\rm sat}$ for simple systems.
Therefore, by observing the saturation regime in a plot of the EP estimator versus $\ell$, we can accurately estimate the total EP (see Sec.~\ref{sec:verification}) without resorting to a time-consuming optimization procedure.

There exists no limitation for choosing a set of $\ell$ weight vectors as long as
they are linearly independent of each other. In this study, we adopt a Gaussian weight vector set~\cite{Vu2020Entropy} for numerical verification of the rate MEB method in Sec.~\ref{sec:verification}. The first weight vector is a Gaussian function, the width of which corresponds to the difference between the maximum and minimum state values. The second and third weight vectors are Gaussian functions with a width half that of the first one, and so on. This represents one way to add the Gaussian basis \emph{uniformly} for a given state-variable range, which yields an accurate and reliable EP estimation as shown in Sec.~\ref{sec:verification}. The mathematical expression for the weight vector set is as follows:
\begin{align}
\{ \Lambda_{i,\alpha}\} = \left\{\cdots, \exp{[-(q_i-a_{i,\alpha})^2/2b^2_{i,\alpha}]}, \cdots \right\}. \label{eq:Gaussian_set}
\end{align}
In Eq.~\eqref{eq:Gaussian_set}, $a_{i,\alpha}$ and $b_{i,\alpha}$ are given as
\begin{align}
    \{a_{i,\alpha}\}_{\alpha \leq \ell} =& \{\frac{1}{2}q_i^{\rm min} + \frac{1}{2}q_i^{\rm max}, \frac{2}{3}q_i^{\rm min} + \frac{1}{3}q_i^{\rm max}, \nonumber\\
    &\,\,\,\frac{1}{3}q_i^{\rm min} + \frac{2}{3}q_i^{\rm max}, \frac{3}{4}q_i^{\rm min} + \frac{1}{4}q_i^{\rm max},... \},\nonumber\\
    \{b_{i,\alpha}\}_{\alpha \leq \ell} =& \{\Delta q_i , \frac{1}{2}\Delta q_i ,\frac{1}{2}\Delta q_i ,\frac{1}{3}\Delta q_i ,\frac{1}{3}\Delta q_i ,\frac{1}{3}\Delta q_i, ... \},\nonumber
\end{align}
where $q_i^{\rm max}$ ($q_i^{\rm min}$) is the maximum (minimum) value of $q_i$ in a given trajectory ensemble and $\Delta q_i = q_i^{\rm max} - q_i^{\rm min}$.
Explicit forms of $a_{i,\alpha}$ and $b_{i,\alpha}$ are given as $a_{i,\alpha} = q_i^{\rm min} + (u(\alpha) + 1)^{-1}[\alpha -\frac{u(\alpha ) (u(\alpha ) - 1)}{2}] \Delta q_i$ and $b_{i,\alpha} = \Delta q_i / u( \alpha ) $ for $\alpha \geq 1 $ with $u(n) = \lfloor \frac{1+\sqrt{8\alpha-7}}{2}\rfloor $. Here, $\lfloor x \rfloor = \tilde m$, where $\tilde m$ is an integer satisfying $\tilde m \leq x <\tilde m+1 $.
We note that the Gaussian weight vector set usually yields reliable estimation results compared to other sets. For example, the polynomial basis set $\{q_i, q_i^2, \cdots , q_i^\ell \}$ typically overestimates the EP with large fluctuations when $\ell$ is large, since the polynomial term with a large exponent is highly sensitive to rare-event data.

\subsection{EP estimation procedure via MEB } \label{sec:EPestimation}

In this section, we describe how to estimate the EP with the MEB from an ensemble of system trajectories. Here we consider both overdamped and underdamped Langevin dynamics. For an overdamped system, the system states consist of only position variables, i.e.,~$\bm q(t) = \bm x(t)= (x_1,\cdots,x_M)$, and the Langevin equation is written as
\begin{align}
	\dot x_i(t) = \frac{1}{\gamma} F_i (\bm q(t),t) + \sqrt{2  B_i (\bm q, t)}\bullet \xi_i (t),
	\label{eq:over_Langevin}
\end{align}
where $F_i$ is a force applied to the $x_i$ component.  The reversible current $J_i^{\rm rev}=0$, while the irreversible current is
given as
\begin{align}
	J_i^{\rm irr} (\bm q, t) = \left( \frac{1}{\gamma} F_i(\bm q, t) -  \partial_{x_i}B_i (\bm q, t)  \right) P(\bm q, t)
	\label{eq:defi_irr_force}.
\end{align}

In the case of underdamped dynamics, the system states consist of both position and velocity variables, i.e.,~$\bm q(t) = (x_1,\cdots,x_N, v_1,\cdots, v_N)$ with $M=2N$, and the Langevin equation is written as
\begin{align}
	&\dot x_i = v_i \nonumber \\
	&\dot v_i(t) = \frac{1}{m} F_i (\bm q(t),t) - \frac{\gamma}{m}v + \sqrt{2  B_i (\bm q, t)}\bullet \xi_i (t).
	\label{eq:under_Langevin}
\end{align}
As done in Eq.~\eqref{eq:rev_irr}, the external force $F_i$ can be divided into reversible and irreversible parts as $F_i = F_i^{\rm rev} + F_i^{\rm irr}$ with
\begin{align}
	F_i^{\rm rev}(\bm q, t) &\equiv  \frac{1}{2} \left[ F_i(\bm q, t) + F_i^\dagger(\bm \epsilon\odot \bm q,t) \right],\nonumber \\
	F_i^{\rm irr}(\bm q, t) &\equiv \frac{1}{2} \left[ F_i (\bm q, t) - F_i^\dagger(\bm \epsilon \odot \bm q,t) \right]. \label{eq:rev_irr_force}
\end{align}
Then, the irreversible currents are given as
\begin{align}
	J_{x_i}^{\rm irr} (\bm q, t) & = 0, \nonumber \\
	J_{v_i}^{\rm irr} (\bm q, t) & = \left( \frac{1}{m} F_i^{\rm irr}(\bm q, t) - \frac{\gamma}{m}v_i - \partial_{v_i} B_i (\bm q, t)  \right) P(\bm q, t)~,  \label{eq:irr_current}
\end{align}
while the reversible currents $J_{x_i}^{\rm rev} (\bm q, t)=v_i P(\bm q, t)$ and
$J_{v_i}^{\rm rev} (\bm q, t)=\frac{1}{m}F_i^{\rm rev}(\bm q, t) P(\bm q, t)$.

We focus on the rate MEB in the following discussions, but note that the procedure for the integral MEB is essentially the same.

\subsubsection{Determination of $ B_i$ and $\textsf L_i$}
\label{sec:Step1}

For an overdamped dynamics described by Eq.~\eqref{eq:over_Langevin}, the diffusivity $ B_i$ is determined from the average of short-time mean square displacements as
\begin{align}
	 B_i (\bm q, t) = \lim_{\delta t \rightarrow 0} \frac{\langle \delta x_i(t)^2 \rangle_{(\bm q, t)}}{\delta t}, ~~\textrm{(overdamped)} \label{eq:B_overdamped}
\end{align}
where $\delta x_i(t) \equiv x_i (t+\delta t) - x_i (t)$ and $\langle \cdots \rangle_{(\bm q, t)}$ denotes  the average over the trajectory ensemble at position $\bm q$ and time $t$.
When $ B_i$ is independent of position and time, all short-time trajectories can be utilized for estimating the diffusivity.
For an underdamped dynamics,  $B_i$ can be estimated from the ensemble of velocity trajectories as
\begin{align}
	B_i (\bm q, t) = \lim_{\delta t \rightarrow 0} \frac{\langle \delta v_i (t)^2 \rangle_{(\bm q, t)}}{\delta t}, ~~\textrm{(underdamped)} \label{eq:B_underdamped}
\end{align}
where $\delta v_i (t) \equiv v_i (t+\delta t) - v_i (t)$.
With these estimations for $B_i$, we calculate
$( \dot{\textsf L}_i^{(\ell)} (t))_{\alpha, \beta}$  from Eq.~\eqref{eq:Lmat_rate}. Note that $\lim_{\delta t \rightarrow 0} \frac{\langle \delta x_i (t)^2 \rangle_{(\bm q, t)}}{\delta t}=0$ in the underdamped case. In the case where the estimation of $B_i (\bm q, t)$ requires heavy computation, instead of evaluating the diffusivity, we can directly estimate the $\left( \dot{\mathsf L}_i^{(\ell)} (t) \right)_{\alpha,\beta}$ matrix from the average of the products of two infinitesimal observable changes as
\begin{align}
	\left( \dot{\mathsf L}_i^{(\ell)} (t) \right)_{\alpha,\beta} &=   \left\langle \Lambda_{i, \alpha} (\bm q)   B_i (\bm q, t)   \Lambda_{i,\beta} (\bm q) \right \rangle_{\bm q} \nonumber \\
	& \approx \langle \Lambda_{i, \alpha} (\bm q)\circ \delta q_i \cdot \Lambda_{i,\beta} (\bm q)   \circ \delta q_i \rangle_{\bm q} /(2\delta t).
\label{eq:obtainL}
\end{align}
When the diffusion matrix has non-zero off-diagonal terms, we first have to directly evaluate ${\mathsf B}_{ij} = \lim_{\delta t\rightarrow 0} \langle \delta q_i \delta q_j \rangle_{(\bm q,t)}/\delta t $ and find the proper transformation to diagonalize the matrix.

\subsubsection{Measurement of an observable current} \label{sec:Step2}

Now we describe how to measure the observable current rate in Eq.~\eqref{eq:Theta_rate} for both overdamped and underdamped dynamics.
First, in the overdamped dynamics with $\bm q(t) = (x_1,\cdots,x_M)$,  the $i$-th component of an observable current rate can be measured by averaging $\Lambda_{i,\alpha} (\bm q (t) )   \circ \dot{x}_i (t)$ over the ensemble of system trajectories as
\begin{align}
	\langle  \dot{\Theta}_{i,\alpha} (t) \rangle = \langle \Lambda_{i,\alpha} (\bm q (t) )   \circ \dot{x}_i (t) \rangle_{(\bm q,t)}, \label{eq:overdamped_current_rate}
\end{align}
where $\circ$ denotes the Stratonovich product. This can be checked from the fact that $\langle H(\bm q, t)\circ \dot{x}_i \rangle_{(\bm q,t)} = \int d \bm q~ H(\bm q, t) J_i^{\rm irr} (\bm q, t)$ with an arbitrary function $H(\bm q, t)$.
For $\Lambda_{i,\alpha} (\bm q) =1$, the observable current is the displacement in the $x_i$ direction as $\Theta_{i,\alpha} (\tau) = x_i (\tau) - x_i (0)$.

In the underdamped dynamics with $\bm q (t) = (x_1, \cdots, x_N, v_1, \cdots, v_N)$, by plugging Eq.~\eqref{eq:irr_current} into Eq.~\eqref{eq:Theta_rate}, we have
\begin{align}
	\langle \dot{\Theta}_{i,\alpha} (t) \rangle  = &  -\frac{\gamma}{m}\langle \Lambda_{i,\alpha} (\bm q) v_i \rangle_{\bm q} + \frac{1}{m} \langle \Lambda_{i,\alpha} (\bm q) F_i^{\rm irr} (\bm q, t) \rangle_{\bm q} \nonumber \\
	& +  B_i (\bm q, t) \langle \partial_{v_i} \Lambda_{i,\alpha} (\bm q )   \rangle_{\bm q}~. \label{eq:observable_underdamped}
\end{align}
When $F_i^{\rm irr}=0$ and $\Lambda_{i,\alpha}$ has no explicit velocity dependence, the observable current is proportional to $\langle \Lambda_{i,\alpha} (\bm q) v_i \rangle_{\bm q}$, similar to that of the overdamped system, Eq.~\eqref{eq:overdamped_current_rate}. However, extra information~$\gamma/m$ is required to evaluate  Eq.~\eqref{eq:observable_underdamped} in the underdamped case. This constant can be determined experimentally by measuring the velocity relaxation time~\cite{JSLee2018}. Otherwise, when $F_i^{\rm irr}=0$ and $\Lambda_{i,\alpha}$ has an explicit velocity dependence, extra calculation of the last term in Eq.~\eqref{eq:observable_underdamped} is necessary.
For the most general case with $F_i^{\rm irr} \neq 0$ (velocity-dependent force),
 $\langle \dot{\Theta}_{i,\alpha} (t) \rangle$ cannot be determined solely by system trajectories, but rather concrete information on the force is necessary.

\subsubsection{EP Estimation}

Utilizing numerical data for  $\dot{\textsf L}_{i}^{(\ell)}$ and $\langle \dot{\Theta}_{i,\alpha} (t)\rangle$ obtained in Secs.~\ref{sec:Step1} and \ref{sec:Step2}, one can evaluate $\sigma^{{\rm MEB}(\ell)}(t)$ from Eq.~\eqref{eq:rateMEB} and then
obtain $\Sigma^{{\rm MEB}(\ell)} (\tau) \equiv \int_0^{\tau} dt~ \sigma^{{\rm MEB}(\ell)} (t)$ for each $\ell = 1, 2, 3, \cdots$.
As proved in Sec.~\ref{sec:MEBderivation}, $\Sigma^{{\rm MEB}(\ell)} (\tau)$ is an increasing function of $\ell$ and saturates to the maximum value at some $\ell^{\rm sat}$. This saturation indicates that $\Lambda_i^{\ell^{\rm sat}}$  coincides with $\Lambda_i^{\rm e} (\bm q, t)$, thus satisfying the equality of the EB. Therefore, the total EP corresponds to the MEB estimator at $\ell=\ell^{\rm sat}$,
i.e.,~$\Delta S^{\rm tot} (\tau) = \Sigma^{{\rm MEB}(\ell^{\rm sat})} (\tau)$.

\section{Relation between MEB and TUR}
\label{sec:comparison}
In this section, we discuss the relation between MEB and TURs. We first consider a one-dimensional (1D) overdamped Langevin dynamics
in the steady state (without a time-dependent protocol) as described by Eq.~\eqref{eq:over_Langevin}.
The total EP $\Delta S^{\rm tot} (t, t^\prime)$ during a small time segment between $t$ and $t^\prime = t+\delta t$ and the corresponding
accumulated current $\Theta (t,t^\prime)$ are written as
\begin{align}
	&\Delta S^{\rm tot} (t, t^\prime) \coloneqq \Delta S^{\rm tot} (t^\prime) - \Delta S^{\rm tot} (t)= \int_t^{t^\prime} dt~ \sigma^{\rm tot} (t), \nonumber \\
	&\Theta (t,t^\prime) \coloneqq \Theta (t^\prime) - \Theta (t) =\int_t^{t^\prime}dt~ \Lambda(x(t) ,t) \circ \dot{x}(t) .
\end{align}
Then, the TUR is given by
\begin{align}
	\Delta S^{\rm tot} (t, t^\prime) \geq \frac{2 \langle \Theta (t, t^\prime) \rangle^2}{\langle \Delta \Theta (t, t^\prime)^2\rangle}, \label{eq:TUR_segment}
\end{align}
where $\langle \Theta (t,t^\prime)\rangle$ and $\Delta \Theta (t, t^\prime )$ are
\begin{align}
	\langle \Theta (t,t^\prime)\rangle & = \int_t^{t^\prime} dt \int d x~\Lambda (x, t) J^{\rm irr} (x, t) ,\nonumber \\
	\Delta \Theta (t, t^\prime ) &= \Theta (t,t^\prime) -  \langle \Theta (t,t^\prime) \rangle .	
\end{align}
In  the short-time limit $\delta t \rightarrow 0$, $\langle \Theta (t,t^\prime) \rangle  = \langle \dot{\Theta} (t)\rangle \delta t$ and  $\langle \Delta \Theta (t,t^\prime)^2\rangle = \langle (\Lambda \circ \delta x - \langle \dot{\Theta} (t)\rangle \delta t)^2 \rangle= 2 \langle \Lambda  B \Lambda \rangle_{x} \delta t + O(\delta t ^{2}) $. With these short-time forms, Eq.~\eqref{eq:TUR_segment} becomes identical to the rate EB equation~\eqref{eq:EB_rate}.
Therefore, the previous EP estimation methods using the multidimensional TUR~\cite{Dechant2019multidimensional,Vu2020Entropy}
are identical to our MEB method for a 1D overdamped Langevin system in the short-time limit.
For a higher-dimensional process, it is not possible to write the component-wise TUR, such as $\Delta S_i(\tau)
\geq 2 \frac{\langle \Theta_i (\tau) \rangle^2}{\langle \Delta \Theta_i (\tau)^2 \rangle}$, and consequently we cannot obtain the component-wise bound for the EP from the TUR. In this sense, the MEB provides more detailed information on the EP, compared to the TUR method.


We note that other modified TURs with an arbitrary time-dependent protocol or an arbitrary initial state do not approach the rate EB in the short-time limit even in one dimension. As an example, consider a 1D overdamped Langevin system driven by a time-dependent protocol. The modified TUR for this process with an arbitrary protocol was introduced by Koyuk and Seifert~\cite{Koyuk2020} as
\begin{align}
	\Delta S^{\rm tot} (t, t^\prime) \geq ~&\frac{2  \left[ \hat{h}(t^\prime) \langle \Theta (t, t^\prime)  \rangle   \right]^2}{\langle \Delta \Theta (t, t^\prime)^2\rangle} \equiv \Delta S^{\rm KS} (t,t^\prime), \label{eq:KSTUR_segment}
\end{align}
where $\hat{h}(t) = t\partial_t - \omega \partial_\omega$ and $\omega$ denotes the protocol speed.
In the $\delta t \rightarrow 0$ limit, the numerator of $\Delta S^{\rm KS}(t,t^\prime)$ becomes $2 \{(1-\omega \partial_\omega) \langle \dot{\Theta} (t) \rangle \}^2 \delta t^2 $, and thus this TUR is written as
\begin{align}
	\sigma^{\rm tot} (t) \geq \frac{2 \{(1-\omega \partial_\omega) \langle \dot{\Theta} (t) \rangle \}^2 }{\langle \Lambda  B  \Lambda \rangle_{x}} \equiv \sigma^{KS} (t), \label{eq:KSTUR_rate}
\end{align}
which is different from the rate EB in general. Note that we have to evaluate the sub-leading-order contribution when the numerator vanishes in Eq.~\eqref{eq:KSTUR_rate}. Experimental estimation of $\sigma^{KS} (t)$ is a very laborious task since we need a sufficiently large ensemble of trajectories, slightly perturbed with respect to $\omega$ at time $t$ for every single $t$. Therefore, compared to this modified TUR method, MEB is a much more efficient approach to correctly estimate the EP of a system driven by a time-dependent protocol.

In addition, short-time TURs for underdamped dynamics do not correspond to the rate EB either. The TUR for a 1D underdamped system with a time-dependent protocol and the observable current $\langle \dot{\Theta} (t) \rangle  =  \langle \Lambda (x, t) v \rangle_{x}$  can be written as~\cite{lee2021universal}  
\begin{align}
	\Delta S^{\rm tot} (t, t^\prime) \geq ~&\frac{2  \left[ \hat{h}_{\rm u}(t^\prime) \langle \Theta (t, t^\prime)  \rangle  \right]^2}{\langle \Delta \Theta (t, t^\prime)^2\rangle} - I(t),  \label{eq:TUR_under}
\end{align}
where $\hat{h}_{\rm u} (t) = t \partial_t - s \partial_s - r \partial_r - \omega \partial_\omega $ and $I(t)$ is the initial-state dependent term, defined as $I(t) = 2 \langle (1+ \hat{h}_{\rm u}^\prime \ln P(x, t))^2 \rangle $ with $\hat{h}_{\rm u}^\prime =  x \partial_{x} - s\partial_s - r \partial_r - \omega \partial_\omega$. Here, $s$ and $r$ are scaling parameters for force and position, respectively.
In the $\delta t \rightarrow 0$ limit, since $\Theta(t,t^\prime) = \Lambda v \delta t$, $\langle \Delta \Theta (t,t^\prime)^2\rangle = \langle (\Lambda  v  - \langle \dot{\Theta} (t)\rangle )^2 \rangle \delta t^2 $, which is not  $O(\delta t)$~\cite{Fischer2020Free}. Thus, in the $\delta t \rightarrow 0$ limit, Eq.~\eqref{eq:TUR_under} becomes
\begin{align}
	\sigma^{\rm tot} (t) \delta t \geq \frac{2 \left[ (1- s \partial_s - r \partial_r- \omega \partial_\omega) \langle \dot{\Theta} (t) \rangle \right]^2 }{\langle (\Lambda  v - \langle \dot{\Theta} (t)\rangle )^2\rangle } - I (t), \label{eq:underTUR_rate}
\end{align}
which is also different from the rate EB. For evaluating Eq.~\eqref{eq:underTUR_rate} experimentally, slight scalings of all forces and position variables are necessary, which demands full knowledge and full controllability of all forces. Thus, it is clear that the underdamped TUR is not practically useful to estimate the EP for a complicated system, such as complex biological systems where such detailed information is not available.

We conclude that the MEB is a unified tool that enables the efficient estimation of the EP from a trajectory ensemble for an overdamped or underdamped Langevin process without an irreversible force. Finally, it is interesting to note that the integral MEB can be tight when we choose the optimal observable current, a feature that no finite-time TUR can achieve.

\section{Numerical Verification of MEB}
\label{sec:verification}

In this section, we estimate the EP of three physical systems driven by time-dependent protocols via the MEB method. All these systems can be realized experimentally using optical tweezers. The first example is a dragged Brownian particle, the second is a pulled harmonic chain, and the last is an RNA unfolding process. We also compare the MEB results to those of other well-established EP measurement methods, namely the direct method utilizing $\Delta S^{\rm tot} = \Delta S^{\rm sys} + Q/T$ and a machine learning technique (NEEP)~\cite{kim2020neep, otsubo2020estimating, kim2021estimating}.

\subsection{Brownian particle dragged by optical tweezers} \label{sec:example1}
\begin{figure}[!t]
	\includegraphics[width=0.48\textwidth]{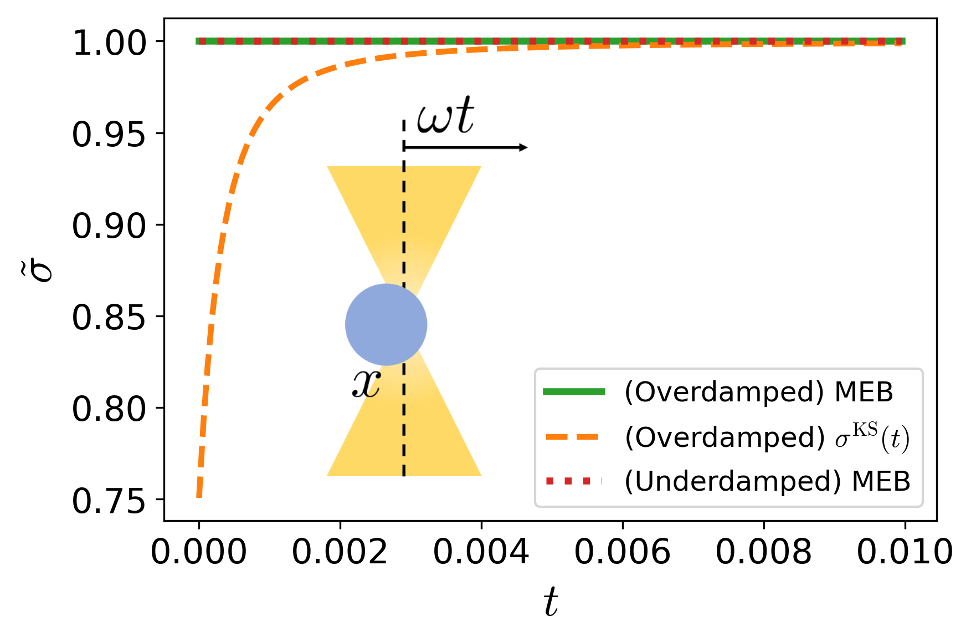}
	\vskip -0.1in
	\caption{ Plot for the rate EP estimator $\tilde\sigma$ normalized with respect to the total EP rate $\sigma^{\rm tot}(t)$ as a function of time $t$. The green solid and red dotted line denote the MEB results of the overdamped and underdamped dynamics, respectively. The orange dashed line represents the result of the modified TUR by Koyuk and Seifert, $\sigma^{\rm KS}(t)/ \sigma^{\rm tot} (t)$. The parameters used for this plot are $k=\mu=\omega=T=1$. The inset shows a schematic of the Brownian particle dragged by optical tweezers. Note that these plots are obtained from the analytic expressions.  
	}
	\label{fig:1dBrown}
	\vskip -0.1in
\end{figure}

We consider a 1D Brownian particle dragged by optical tweezers as illustrated in Fig.~\ref{fig:1dBrown}. The center of the harmonic potential of the tweezers is initially ($t \leq 0$) located at the origin and moves with a constant speed $\omega$ for $t>0$. Then the corresponding overdamped Langevin equation  for the position $x(t)$ is written as
\begin{align}
    \dot x (t) = - \mu k  (x(t) - \lambda(t) ) + \sqrt{2 B }\xi(t)
    \label{eq:1D_Brown}
,\end{align}
where $\lambda(t)= \omega t$ is a  time-dependent protocol, $\mu$ is the mobility, $k$ is the spring constant of the harmonic potential, and $B = \mu T$ with the environmental temperature $T$.
The initial state at $t=0$ is set as the equilibrium state. As the driving force is linear in position, we can solve the equation of motion analytically. The procedure for deriving the analytic solutions is presented in Appendix~\ref{sec:analytic_1D_dragged}.

\begin{figure*}[!t]
	\includegraphics[width=\textwidth]{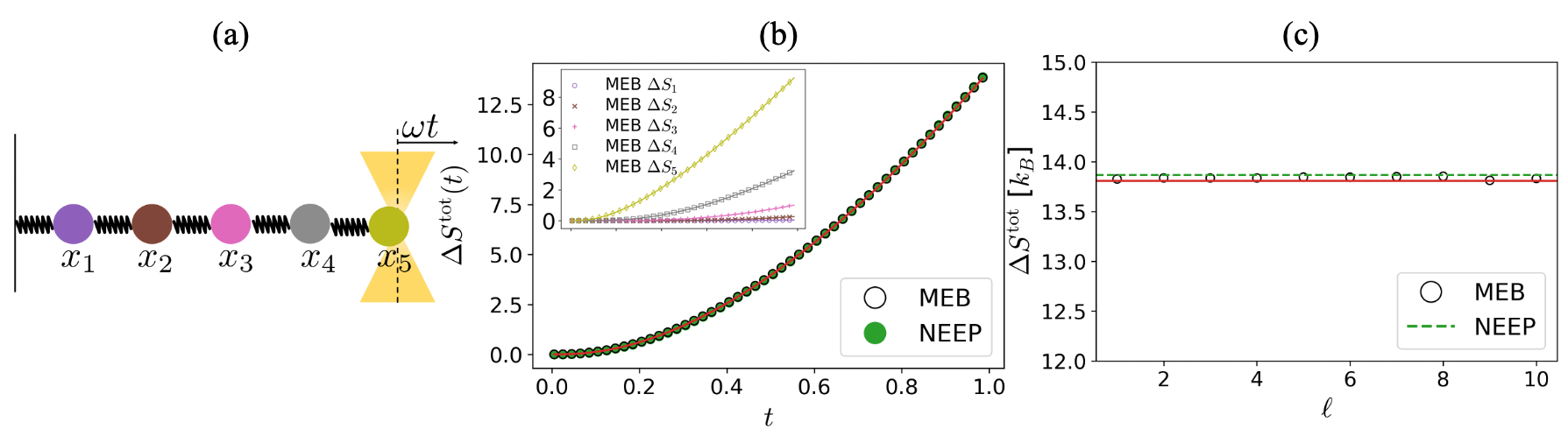}
	\vskip -0.1in
	\caption{
		(a) Schematic of the harmonic chain pulled by optical tweezers. (b) Estimated EP via the MEB method (black) and the NEEP method (green) as a function of time $t$. The inset shows the EP of the $i$-th bead. $\bigcirc$, $\times$,  $+$, $\square$, and $\Diamond$ represent the estimated EPs for $x_1$, $x_2$, $x_3$, $x_4$, and $x_5$ beads, respectively. Solid lines denote the analytic results. Four Gaussian weight vectors are used for the MEB estimation. (c) Plot of $\Delta S^{\rm tot}$ at $t=1.0$ against the number of weight vectors~$\ell$. The green dashed line represents the NEEP result, while the red solid lines in (b) and (c) denote the analytic results. The parameters used for these plots are $k=5$, $\mu =1$, $\omega = 5$, and $T =1$.
	}
	\label{fig:5beads}
	\vskip -0.1in
\end{figure*}

We measure the displacement of the particle, that is, we take the weight vector $\Lambda(x,t) = 1$. With this observable current, we evaluate the rate MEB estimator $\sigma^{\rm MEB}(t)$ for each time $t$ analytically and plot the results in Fig.~\ref{fig:1dBrown}. Note that the normalized estimator is defined by $\tilde{\sigma}  \equiv \sigma^{\rm MEB}(t)/\sigma^{\rm tot}(t)$, which turns out to be unity, i.e.,~the estimated EP exactly matches the true EP.
This is a rather surprising result, as we use only one current (displacement).
In fact, one can analytically find the tight weight factor $\Lambda^{\rm e}(x,t)$ in Eq.~\eqref{eq:weight_rate} with
the help of the exact solution in Eq.~\eqref{eqA:Jirr_Brownian} as
\begin{align}
\Lambda^{\rm e} (x,t) = c(t) \frac{J^{\rm irr}(x,t)}{B P(x,t)} =c(t) \frac{\omega}{\mu T} (1-e^{-t/\tau_\mu}), \label{eq:tight_condition_single_Brownian}
\end{align}
where $\tau_\mu = 1/(\mu k) $. Note that $\Lambda^{\rm e}(x,t)$ is $x$-independent. Thus, by choosing
the arbitrary $c(t)$ to cancel the $t$-dependence exactly in Eq.~\eqref{eq:tight_condition_single_Brownian},
one can easily see that the unity weight vector $\Lambda^{\rm e}= 1$ also satisfies the equality condition of the rate EB.

For the purpose of comparison, we also plot the ratio of the modified TUR by Koyuk and Seifert~\cite{Koyuk2020} to the EP rate
in Fig.~\ref{fig:1dBrown}. 
For evaluating the ratio of this example, the sub-leading-order terms that we neglected for deriving Eq.~\eqref{eq:KSTUR_rate} are necessary since the numerator in Eq.~\eqref{eq:KSTUR_rate} vanishes in this example, so we calculate $\Delta S^{KS}(t+\delta t,t)/\Delta S(t+\delta t,t)$ with small $\delta t =0.001$. This approximated ratio coincides with the result in Ref.~\cite{Koyuk2020}.
The modified TUR deviates largely from the correct one for small $t$.
 This confirms that our MEB method outperforms the modified TUR method for this simple case.

We also consider the same process in the underdamped version. The corresponding underdamped Langevin equation is written as
\begin{align}
        \dot x (t) =& v (t) \nonumber\\
        \dot v (t)  =& - \frac{1}{m \mu} v (t) -\frac{k}{m} (x(t) - \lambda (t) ) + \sqrt{2 B  } \xi(t),
        \label{eq:under_dragged}
\end{align}
where $\lambda(t) = \omega t$ and $B = T/(\mu m^2)$. The initial state is also set as the equilibrium state. The analytic solution of this equation is also available via similar procedure for solving Eq.~\eqref{eq:1D_Brown}. The derivation is presented in Appendix~\ref{sec:analytic_1D_dragged}. From Eq.~\eqref{eqA:underdamped_irr_current}, the tight weight vector is obtained as
\begin{align}
    \bm\Lambda^{\rm e}(x,v,t) = c(t) \frac{\bm J^{\rm irr}(x,v,t)}{B P(x,v,t)} = c(t)
    \begin{pmatrix}
        0  \\  -  \frac{m \langle v(t)\rangle}{T}
    \end{pmatrix} ,
    \label{eq:tight_condition_under_Brownian}
\end{align}
where $\langle v(t) \rangle $ is evaluated by taking the time derivative of $\langle x(t)\rangle$ in Eq.~\eqref{eqA:underdamed_x_avg_sol}.
We find that $\bm \Lambda^{\rm e}(x,v,t)$ depends only on time but not position, like in the overdamped case. Therefore, the unit weight vector again provides the EP exactly. The analytic result of $\tilde{\sigma}  = \sigma^{\rm MEB}(t)/\sigma^{\rm tot}(t)$ for this underdamped dynamics is also plotted in Fig.~\ref{fig:1dBrown}, which confirms the exact estimation of the EP from the rate MEB by measuring only the displacement.

\subsection{Harmonic chain pulled by optical tweezers}

The next example is an $M$-bead harmonic chain dragged by optical tweezers as illustrated in Fig.~\ref{fig:5beads}(a). The harmonic potential of the optical tweezers is exerted on the rightmost particle of the chain, and the leftmost spring clings to the wall. Here, we consider an overdamped Langevin dynamics described by
\begin{align}
    \dot{ \bm  x }(t) = - \mu \mathsf K \bm x(t) + \mu k \bm \lambda(t) + \sqrt{2 \mathsf B } \bm \xi (t),
    \label{eq:Langevinfor5beads}
\end{align}
where $K_{ij} = 2k\delta_{i,j} -k(\delta_{i+1,j} +\delta_{i-1,j} )$, $ \lambda_i = \omega t \delta_{M,i} $, and $ B_{ij} = \mu T \delta_{i,j}$ with $i,j \in \{1, ..., M\}$. We can solve Eq.~\eqref{eq:Langevinfor5beads} in a similar way used for the dragged Brownian particle. The derivation details are presented in Appendix~\ref{sec:analytic_1D_dragged}.

\begin{figure*}[!t]
\centering
\includegraphics[width=\textwidth]{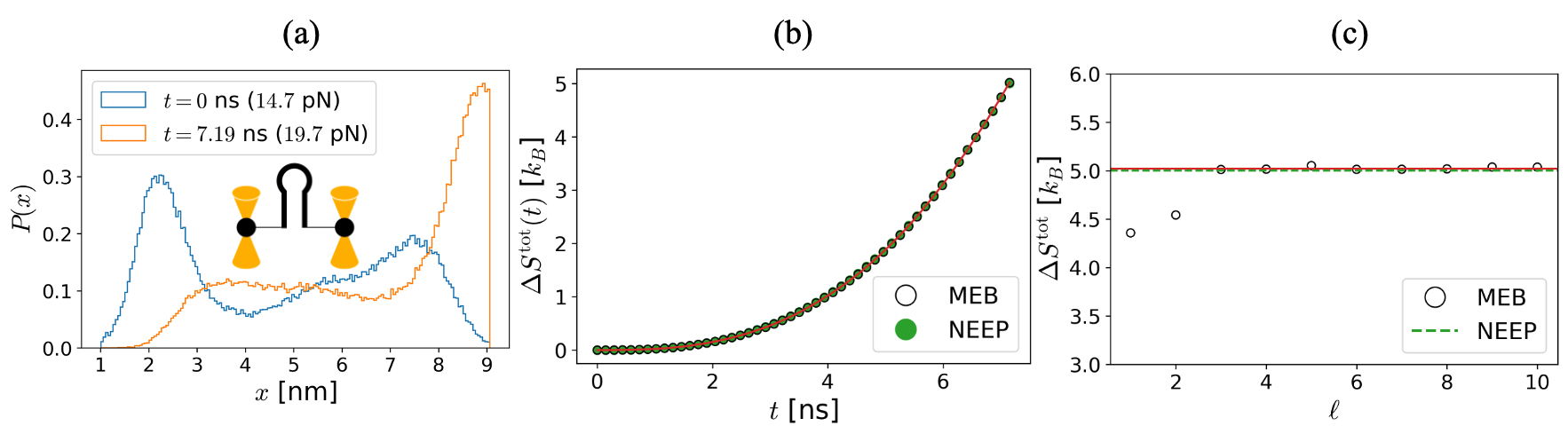}
\vskip -0.1in
\caption{ EP estimation results for the RNA unfolding process.
(a) Histogram of the distance~$x$ between the two ends of P5GA at initial time~$0\text{ ns}$ (light blue) and final time~$7.19\text{ ns}$ (orange). The inset shows a schematic of the RNA pulled by optical tweezers.
(b) EP estimated via MEB with four Gaussian weight vectors~(black) and NEEP~(green) as a function of time~$t$.  The red solid line denotes the results obtained from Eq.~\eqref{eq:tot_ent_def}.
(c) Estimated EP via MEB at $t=7.19\,\text{ns}$ as a function of the number of weight vectors~$\ell$ (black). The green dashed line and the red solid line denote the results of the NEEP and the EP obtained from Eq.~\eqref{eq:tot_ent_def}, respectively.}
\label{fig:RNA}
\vskip -0.1in
\end{figure*}

For validating the MEB estimator, we generate $10^5$ trajectories of $\tau = 1$ by solving Eq.~\eqref{eq:Langevinfor5beads} numerically with $M=5$ and $T=1$ via the 2nd-order stochastic differential equation integrator. We set the time resolution~$dt$ to $0.01$. The initial state of the chain is in equilibrium, with the center of the harmonic potential being located at the origin. From the trajectories, we estimate $\Delta S_i$, the EP for the $i$-th bead, by using the rate MEB estimator with the $\ell=4$ Gaussian weight vector set. The estimated data are plotted in the inset of Fig.~\ref{fig:5beads}(b). As the figure shows, the estimated EP of each bead perfectly matches the analytic result. We plot the total EP by adding all these $\Delta S_i$ in Fig.~\ref{fig:5beads}(b).
For comparison, we also estimate the total EP with the NEEP machine learning technique~\cite{kim2020neep}, which coincides with the MEB result precisely. The detailed procedure for the NEEP calculation is explained in Appendix~\ref{appendix:NEEP}. Both methods exactly estimate the total EP within $0.5\%$ error.

Figure~\ref{fig:5beads}(c) is a plot of the total EP at $t=1$ against the number of weight vectors. Surprisingly, the EP estimated by the MEB with only one weight vector is already very close to the analytic result. This is due to the fact that a constant weight vector results in the exact EP value in this system, as explained Appendix~\ref{sec:analytic_1D_dragged}. As a Gaussian function with a broad width can be approximated as a constant, the EP can be approximately estimated solely with the broadest Gaussian function. The EP for $\ell >1$  saturates to the analytic result as expected in Sec.~\ref{sec:MEBderivation}.
In Fig.~\ref{fig:5beads}(c), we also plot the result of the NEEP calculation, which is also close to the analytic result.

To test the performance of the MEB method for a high-dimensional system, we perform the same simulation with $M=100$. The estimated EP from the MEB method is $ 14.4$, which is only $3.5\%$ distant from the analytic result, $13.9$. We note that this accurate estimation is infeasible for the direct (plug-in) method for such a high-dimensional system.

\subsection{RNA unfolding process}
\label{sec:RNA}

The final example is an RNA unfolding process, which involves a nonlinear potential and thus an analytic treatment is not possible. A typical experimental setup consists of a single RNA hairpin molecule whose terminals are connected to DNA handles that are controlled by two optical tweezers, as illustrated in Fig.~\ref{fig:RNA}(a). By moving the center of the rightmost optical tweezers, a pulling force is exerted on the rightmost particle and the RNA is unfolded. For the RNA hairpin P5GA~\cite{KimPNAS2012}, a pulling force that amounts to $14.7$~pN yields equal probabilities for folded and unfolded states. The governing equation of motion in this case is
\begin{align}
    \dot x(t) = f_{14.7}(x(t)) + \omega t + \sqrt{2D} \xi(t),
\end{align}
where $x(t)$ is the distance between the two ends of the RNA at time~$t$. The force function $f_{14.7}$ is estimated from coarse-grained molecular dynamics simulation data  when the RNA is pulled by a $14.7$ pN force~\cite{KimPNAS2012}, where a polynomial function of degree~$10$ is employed to fit the force. The reflection boundary condition is imposed at $x_{\rm min} = 1.01$~nm and $x_{\rm max } = 9.07\,\text{nm} $, as distances larger than $x_{\rm max} $ and smaller than $x_{\rm min}$ were not found in the simulation~\cite{KimPNAS2012}. We set the initial condition as the equilibrium state at room temperature~$300$~K, which is an ordinary experimental setup. During the process time $\tau = 7.19~\text{ns}$, the pulling force linearly increases up to $19.7\,\text{pN}$ with a constant rate $\omega=\frac{5.0}{7.19}$~pN/ns. We generate $10^5$ trajectories from the simulations. The time resolution~$dt$ is set to 79.1~ps. The initial and  final distributions at $t=0$ and $\tau$ are presented in Fig.~\ref{fig:RNA}(a), respectively.

We estimate the total EP by evaluating the rate MEB from the trajectory ensemble. Here we use the Gaussian weight vector set from Eq.~\eqref{eq:Gaussian_set} for evaluating the rate MEB estimator. Figure~\ref{fig:RNA}(b) shows a plot of the estimated EP as a function of time for $\ell = 4$. As the analytic expression of the EP for this system is not available, we evaluate the EP using other numerical methods to check the validity of the MEB method. First, we employ the NEEP using the same trajectory ensemble and present the result in Fig. 3(b). We find that the MEB and NEEP results coincide with each other precisely. Second, we apply the direct method using the following equality:
\begin{align}
    \Delta S^{\rm tot } (\tau) =   \Delta S^{\rm sys} (\tau) + \frac{1}{T}\int_0^{\tau} \left\langle (f_{14.7}(x) +\omega t )\circ \dot x  \right\rangle dt ,
    \label{eq:tot_ent_def}
\end{align} where
$\Delta S^{\rm sys } (\tau) = \langle -\ln{p(x,\tau) + \ln{p(x,0)} } \rangle $.  The integration over the process time of the last term in Eq.~\eqref{eq:tot_ent_def} denotes the dissipated heat during the process. The initial and the final PDFs can be estimated from the trajectory ensemble. This task becomes much harder with increasing system dimension. The estimated total EP from the direct method is denoted as the red solid line in Fig.~\ref{fig:RNA}(b), which matches the MEB and NEEP results very well.
We note that the computational cost of the MEB method is lower than that of the NEEP method; it takes 3 s for the MEB method with $4$ weight vectors, while it takes 60 s for the NEEP process including the learning time~\footnote{Intel i9-11900K is used for MEB evaluation. The same CPU and an RTX 3090 are used for NEEP calculation.}. Here, we do not take into account the time required to find the proper hyperparameters for the NEEP.

Figure~\ref{fig:RNA}(c) shows the way to determine the proper number of weight vectors $\ell$. From a given trajectory ensemble, we estimate the total EP by using the MEB estimator; the estimated EP as a function of $\ell$ for this RNA unfolding process is plotted in Fig.~\ref{fig:RNA}(c). For $\ell < 3$, the estimated EP increases as $\ell$ increases, which indicates that no combinations of two Gaussian weight vectors, Eq.~\eqref{eq:linearComb_rate}, are sufficiently close to the optimal weight vector, Eq.~\eqref{eq:weight_rate}. The estimated EP saturates to a certain value for $\ell\geq 3 $, which indicates that the estimator is now sufficiently close to the optimal one. Thus, accurate EP estimation can be obtained by choosing $\ell\geq 3$. We also examine the dependence of the time resolution of an experiment and a limited number of trajectories on the EP in Appendix~\ref{appendix:RNAlimiteddata}.

\section{Conclusion and Discussion}
\label{sec:disscussion}

In this study, we suggested an EP estimator, named MEB, by applying multidimensional observable currents to the entropic bound. The MEB provides a unified way to estimate the EP for both overdamped and underdamped Langevin dynamics regardless of the time dependence of the protocol. The MEB estimator can be obtained in either integral or rate form. The tight EP bound is always achievable for any finite-time processes via both the integral and the rate MEBs, whereas it is possible for TURs only in the short-time limit. From numerical simulations, we confirmed that the MEB estimates the EP with high accuracy from an ensemble of system trajectories of overdamped Langevin systems. For an underdamped system with an irreversible force, information about the force is additionally required to estimate the EP. Moreover, extra information on the relaxation time is necessary for underdamped systems. Therefore, a precise estimation of the EP may be possible via MEB even for various complicated physical processes, in particular biological systems.

In future research, it will be interesting to develop a method to estimate the stochastic EP at the level of a single trajectory for general Langevin dynamics, rather than the averaged EP over an ensemble of trajectories.
Moreover, the extension of the EP estimation to an open quantum system will be another intriguing problem. The quantum TUR recently proposed in Ref.~\cite{Van2022Thermodynamics} could be a good candidate for an estimator of the EP of an open quantum system, if one can measure the coherent-effect term in the formulation.
It is also worthwhile to mention the recently proposed method for directly inferring the stochastic differential equations from a given trajectory ensemble~\cite{Bruckner2020Inferring,Frishman2020learning}. It would be interesting to compare the accuracy and efficiency of EP estimation between MEB and the inferring method.

\begin{acknowledgments}
This research was supported by an NRF Grant No. 2017R1D1A1B06035497 (H.P.) and KIAS Individual Grants Nos. PG081801 (S.L.), PG074002 (J.-M.P.), CG076002 (W.K.K.), QP013601 (H.P.), and PG064901 (J.S.L.) at the Korea Institute for Advanced Study. D.-K.K. was supported by the Institute for Basic Science of Korea (IBS-R029-C2).
\end{acknowledgments}

\appendix

\section{Derivation of Eq.~\eqref{eq:increasing_l}}
\label{appendix:MEBandaddingbasis}

Here, we focus on the derivation of the integral MEB. The derivation for the rate MEB is essentially the same as that of the integral MEB. ${\mathsf L}_i^{(\ell+1)}$ can be expressed as the following block matrix form:
\begin{align}
    {\mathsf L}_i^{(\ell+1) } =
    \begin{bmatrix}
        {\mathsf L}_i^{(\ell)} & \bm b\\
        \bm b^{\mathsf T} & h \label{eqA:block_L}
    \end{bmatrix}
\end{align}
where $\bm b^{\mathsf T} = [ ({\mathsf L}_i)_{\ell+1,1}, \cdots , ({\mathsf L}_i)_{\ell+1,\ell} ] $ and $ h = ({\mathsf L}_i)_{\ell+1, \ell+1} = \int^\tau_0 dt \langle  \Lambda_{i,\ell+1}   \mathsf B_i  \Lambda_{i, \ell+1}  \rangle_{\bm q}  $. From the Schur complement, the determinant of the block matrix $L_i^{(\ell +1)}$ in Eq.~\eqref{eqA:block_L} is given by
\begin{align}
	{\rm{det}} ({\mathsf L}_i^{(\ell+1)} ) = {\rm{ det }} ( {\mathsf L}_i^{(\ell)}) \left[ h - \bm b^{\mathsf T} ({\mathsf L}_i^{(\ell)} )^{-1} \bm b \right].
	\label{eqA:det}\end{align}
The determinant of $\mathsf
L_i^{(\ell)}$ for any $\ell$ is positive since it is a positive-definite matrix. This implies that the term $h - \bm b^{\mathsf T} ({\mathsf L}^{(l)} )^{-1} \bm b $ in Eq.~\eqref{eqA:det} is also positive. Moreover, via the following inverse block matrix formula,
\begin{widetext}
\begin{align}
	\begin{bmatrix}
		\mathsf A & \mathsf B \\
		\mathsf C & \mathsf D
	\end{bmatrix}^{-1}
	=&   \begin{bmatrix}
			\mathsf A^{-1}+ \mathsf A^{-1} \mathsf B (\mathsf D - \mathsf C \mathsf A^{-1}\mathsf B)^{-1} \mathsf C \mathsf A^{-1}  & -\mathsf A^{-1} \mathsf B (\mathsf D - \mathsf C \mathsf A^{-1} \mathsf B )^{-1} \\
		-(\mathsf D - \mathsf C \mathsf A^{-1} \mathsf B )^{-1} \mathsf C \mathsf A^{-1}   & (\mathsf D - \mathsf C \mathsf A^{-1} \mathsf B )^{-1}
	\end{bmatrix}
	,\end{align}
\end{widetext}
the inverse matrix of $\mathsf L_i^{(\ell+1)}$ can be expressed as
\begin{align}
    ({\mathsf L}_i^{(\ell+1)})^{-1}
    =& \begin{bmatrix}
        ({\mathsf L}_i^{(\ell)})^{ - 1 } & 0 \\
        0 & 0
    \end{bmatrix}
    + (h -\bm b^{\mathsf T} ({\mathsf L}_i^{(\ell)})^{-1} \bm b )^{ - 1 } \bm d    \bm d^{\mathsf T} ,
    \label{eqA:L_inverse}
\end{align}
where $\bm d^{\mathsf T}  \equiv ( - \bm b^{\mathsf T} ({\mathsf L}_i^{(\ell)})^{ - 1 } , 1)$. Using Eq.~\eqref{eqA:L_inverse}, we can prove Eq.~\eqref{eq:increasing_l} as follows:
\begin{align}
	\Delta  S_i^{{\rm MEB} (\ell+1) } =& \langle {\bm \Theta}_i^{(\ell+1)}\rangle^{\mathsf T}  ({\mathsf L}_i^{(\ell+1) })^{-1}  \langle {\bm \Theta}_i^{(\ell+1)} \rangle \nonumber \\
	=& \Delta  S_i^{{\rm MEB} (\ell)} + ( h - {\bm b}^{\mathsf T} ({\mathsf L}_i^{(\ell)})^{-1} {\bm b} ) (\bm d^{\mathsf T} \langle \bm \Theta_i^{(\ell+1)} \rangle)^2 \nonumber \\
	\geq& \Delta  S_i^{{\rm MEB}(\ell)}. \label{eqA:increasingEP}
\end{align}
The positiveness of $h - {\bm b}^{\mathsf T} ({\mathsf L}_i^{(\ell)})^{-1} {\bm b}$ is used for showing the last inequality of Eq.~\eqref{eqA:increasingEP}.

\section{Analytic solutions of a dragged Brownian particle and pulled harmonic chain by optical tweezers }
\label{sec:analytic_1D_dragged}
We consider a Brownian particle dragged by optical tweezers of which dynamics is governed by the following equation~\cite{Koyuk2020, Van2003Stationary}:
\begin{align}
    \dot x (t) = - \mu k  (x(t) - \lambda (t) ) + \sqrt{2 B }\xi(t),
\label{eq:Gen_drag}\end{align}
where $B=\mu T$ and $\lambda (t)$ is an arbitrary time-dependent protocol with the condition $\lambda(0)=0$. The initial state is set as the equilibrium distribution of Eq.~\eqref{eq:Gen_drag} with $\lambda(0)=0$, and thus, $\langle x(0)\rangle =0$.
To obtain the analytic solution of Eq.~\eqref{eq:Gen_drag}, we decompose $x(t)$ into the deterministic part~$\langle x(t) \rangle$ and the stochastic part~$X(t)\equiv x(t) - \langle x(t) \rangle$.
Taking the average of both sides of Eq.~\eqref{eq:Gen_drag} leads to an equation for the deterministic part as
\begin{align}
    \langle \dot x (t) \rangle = - \mu k (\langle x(t) \rangle - \lambda(t))
.\end{align}
Then, the solution of $\langle x(t)\rangle $ is given by
\begin{align}
    \langle x(t) \rangle = &e^{-t/\tau_{\mu}} \langle x(0) \rangle + \tau_{\mu}^{-1} \int^{t}_{0} dt' e^{-(t-t')/\tau_{\mu}} \lambda(t')\nonumber\\
    =& \lambda(t) - \int^{t}_{0} dt' e^{-(t-t')/\tau_{\mu}} \dot{\lambda}(t')
    \label{eq:1d_Brown_x_ave}
,\end{align}
where $\tau_{\mu} = (\mu k)^{-1} $ is a characteristic  relaxation time. The equation for the stochastic component~$X(t)$ can be obtained by simply substituting $X(t) + \langle x(t) \rangle$ for $x(t)$ in Eq.~\eqref{eq:Gen_drag} as
\begin{align}
    \dot X(t) = -\tau_{\mu}^{-1} X(t) + \sqrt{2B}\xi(t).
\end{align}
Since the initial state is in equilibrium, the distribution of $X(t)$ does not change in time. Therefore, the distribution for all time is given by the equilibrium distribution as
\begin{align}
    P(X,t) = \sqrt{\frac{\beta k }{2 \pi}} \, e^{- \frac{k}{2}\beta X^2 }. \label{eqA:equi_dist}
\end{align}
By substituting $x-\langle x(t) \rangle$ for $X$ in Eq.~\eqref{eqA:equi_dist}, we have
\begin{align}
    P(x,t) = \sqrt{\frac{\beta k }{2 \pi}} \, e^{- \frac{k}{2}\beta (x-\langle x(t)\rangle)^2 }.  \label{eqA:Pxt}
\end{align}
Using Eqs.~\eqref{eq:Jirr} and \eqref{eq:1d_Brown_x_ave}, the irreversible current is \begin{align}
    J^{\rm irr}(x,t) =& \left[-\mu k (x - \lambda(t) ) - B\partial_x \right] P(x,t)\nonumber\\
    =& \mu k [\lambda(t) - \langle x(t) \rangle ] P(x,t)\nonumber \\
    =& \mu k \int^{t}_{0} dt' e^{-(t-t')/\tau_{\mu}} \dot \lambda(t') P(x,t).
\end{align}
When $\lambda(t)=\omega t$ as in  Sec.~\ref{sec:example1}, the irreversible current and EP rate are
\begin{align}
    J^{\rm irr}(x,t) =& \omega [ 1 - e^{- t / \tau_\mu } ] P(x, t), \label{eqA:Jirr_Brownian}  \\
   \sigma^{\rm tot}(t) =& \int dx \frac{  {J}^{\rm irr} (x, t)^2 }{BP(x, t)}\nonumber\\
   =& \frac{\omega^2 \beta}{\mu} (1 - e^{-t /\tau_{\mu}})^2
.\end{align}

The derivation procedure for the underdamped Langevin equation~\eqref{eq:under_dragged} is essentially the same as that of the overdamped equation. By decomposing $x(t)$ into $\langle x(t)\rangle$ and $X(t) = x(t) - \langle x(t)\rangle$ and $v(t)$ into $\langle v(t)\rangle$ and $V(t) = v(t) - \langle x(t)\rangle$, we have
\begin{align}
	\frac{d^2}{dt^2}\langle x(t)\rangle &= -\frac{1}{m \mu} \frac{d}{dt} \langle x(t) \rangle - \frac{k}{m} (\langle x(t) \rangle - \omega t ), \label{eqA:underdamped_x_avg} \\
	\frac{d}{dt} V(t) &= - \frac{1}{m \mu} V(t) - \frac{k}{m} X(t) + \sqrt{2B} \xi (t). \label{eqA:underdamped_V}
\end{align} For this underdamped case, $B = T\mu^{-1}m^{-2}$.
The second-order differential equation~\eqref{eqA:underdamped_x_avg} can be solved  with the boundary conditions $\langle x(0) \rangle = 0$ and $\langle v(0) \rangle = d/dt \langle x(t) \rangle  |_{t=0} =0$. The result is
\begin{align}
	\langle x(t) \rangle = C_+ e^{\alpha_+ t} + C_- e^{\alpha_- t} + \omega t - \frac{\omega}{\mu k}, \label{eqA:underdamed_x_avg_sol}
\end{align}
where $\alpha_{\pm} = -1/(2 m\mu) \pm \sqrt{1/(2 m\mu)^2 - k/m}$ and $C_\pm = \mp \omega(\alpha_\mp/(\mu k)+1) /(\alpha_+ - \alpha_- )$.
As the initial state is in equilibrium, the distribution of $X(t)$ and $V(t)$ in Eq.~\eqref{eqA:underdamped_V} for all time is the following equilibrium distribution:
\begin{align}
	P(X,V,t) = \sqrt{\frac{\beta k}{2 \pi} } \sqrt{\frac{\beta m}{2 \pi} } \exp\left[ -\frac{\beta}{2} (k X^2 + m V^2) \right]. \label{eqA:PXVt}
\end{align}
Therefore, $P(x,v,t)$ is given by substituting $x-\langle x(t) \rangle$ for $X$ and $v-\langle v(t) \rangle$ for $V$ in Eq.~\eqref{eqA:PXVt}, as was done in Eq.~\eqref{eqA:Pxt}.
From Eq.~\eqref{eq:Jirr}, the irreversible current is written as
\begin{align}
	J_x^{\rm irr}(x,v,t) = 0,~~~ J_v^{\rm irr}(x,v,t) = - \frac{\langle v(t) \rangle}{m\mu} P(x,v,t). \label{eqA:underdamped_irr_current}
\end{align}
Finally, the EP rate is evaluated as
\begin{align}
	\sigma^{\rm tot}(t) = \int dx \frac{  {J}^{\rm irr} (x, v,t)^2 }{BP(x, v, t)}	= \frac{\langle v (t)\rangle^2}{\mu T}.
\end{align}

The analytic solution of Eq.~\eqref{eq:Langevinfor5beads} can be obtained in a similar way. By decomposing $x_i (t)$ into $X_i (t) = x_i (t) - \langle x_i (t) \rangle$ and rearranging the terms of Eq.~\eqref{eq:Langevinfor5beads}, we have
\begin{align}
	    \langle \dot{ \bm  x }(t) \rangle &= - \mu \mathsf K \bm \langle \bm x(t) \rangle + \mu k \bm \lambda(t),  \label{eqA:HarmonicChain1} \\
	    \dot{ \bm  X }(t) &= - \mu \mathsf K \bm X(t) + \sqrt{2 \mathsf B } \bm \xi (t) . \label{eqA:HarmonicChain2}
\end{align}
The expression of $\langle x_i (t) \rangle $ can be obtained by solving Eq.~\eqref{eqA:HarmonicChain1}, and it is certain that $\langle x_i (t) \rangle $ is a function of time. Since the initial state is the equilibrium state of Eq.~\eqref{eqA:HarmonicChain2}, the probability density function (PDF) is given by the Boltzmann factor $\exp[-\beta U(\bm X) ]$, where $U(\bm X) =\frac{1}{2} \bm X^{\textsf T} \textsf K \bm X$ is the potential energy of the harmonic chain. Thus, by substituting $\bm X = \bm x - \langle \bm x\rangle $ into the Boltzmann factor, the PDF is written as
\begin{align}
	P(\bm x,t) =&  \frac{e^{- \frac{\beta}{2} (\bm x - \langle \bm x(t)\rangle )^{\mathsf T}  \mathsf K  ({\bm x} - \langle \bm x(t)\rangle)  }}{\sqrt{\det(2 \pi \textsf K^{-1}/\beta  ) } } ,\\
	\bm J^{\rm irr} (\bm x, t) =& [ \mu k \bm \lambda (t) - \mu \mathsf K   \langle \bm x(t)  \rangle   ]P(\bm x, t).
\end{align}
The tight weight vector is then given by
\begin{align}
	\Lambda_i^{\rm e}(\bm x, t) &= c_i(t) \frac{J_i^{\rm irr}(\bm x, t )}{ \mu T P(\bm x, t)} \nonumber \\
	&= c_i (t) \left(\mu k \lambda_i (t) - \mu \sum_j K_{ij} \langle x_j (t) \rangle \right).
\end{align}
Note that $J_i^{\rm irr}(\bm x, t ) / P(\bm x, t) $ depends on time but not position. Thus, the MEB estimator evaluated by measuring displacement, i.e., $\Lambda_i^{\rm e}(\bm x, t)=1$, results in the correct EP.

\section{Component-combined MEB}
\label{appendix:non-component-wse MEB}
In this section, we present the derivation of the component-combined MEB, which is useful when the diffusion matrix has off-diagonal elements. To this end, similar to Eq.~\eqref{eq:linearComb},  we consider the following linear combination of weight vectors as
\begin{align}
   \bm \Lambda^{(\ell)}(\bm q, t) = \sum_{\alpha=1}^{\ell} k_\alpha \bm \Lambda_\alpha (\bm q, t).
\label{eq:linearCombNCW}
\end{align}
After substituting $\bm \Lambda$ in Eq.~\eqref{eq:EB_integral1} with $\bm \Lambda^{(\ell)}(\bm q, t)$ in Eq.~\eqref{eq:linearCombNCW}, we have
\begin{align}
    \Delta S^{\rm tot} (t) \geq& \frac{\left [\int^\tau_0 dt \int d \bm q \bm \Lambda^{(\ell)}(\bm q,t)^{\textsf T} \bm J^{\rm irr} (\bm q,t )  \right ]^2}{ \int^\tau_0 dt \langle \bm \Lambda^{(\ell)}(\bm q,t)^{\textsf T} \mathsf B  \bm \Lambda ^{(\ell)}(\bm q,t) \rangle_{\bm q}  }   \nonumber\\
    =& \frac{ (\bm k^{\textsf T}  \langle\bm \Theta^{(\ell)} (\tau)\rangle )^2 }{\bm k^{\textsf T} \mathsf L^{(\ell)} (\tau)  \bm k } = \Delta {\hat S}^{(\ell)}(\bm k), \label{eqA:MEB_combined1}
\end{align}
where the components of $\langle\bm \Theta^{(\ell)} (\tau)\rangle$ and $\mathsf L^{(\ell)} (\tau)$ are given as
\begin{align}
    \langle \Theta^{(\ell)}_\alpha (\tau) \rangle =&\int^\tau_0 dt \int d \bm q \bm \Lambda_{\alpha}(\bm q,t)^{\textsf T} \bm J^{\rm irr} (\bm q,t ),  \\
     (\mathsf L^{(\ell)} (\tau ))_{\alpha,\beta} =& \int^\tau_0 dt \langle \bm \Lambda_\alpha(\bm q,t)^{\textsf T}  \mathsf B  \bm \Lambda_\beta(\bm q,t)  \rangle_{\bm q}.
\end{align}
Note that $\mathsf L^{(\ell)} (\tau)$ is a positive definite matrix since $\bm z  \mathsf L^{(\ell)}(\tau)  \bm z =  \int^\tau_0 dt  \langle ||\sum_\alpha \mathsf B^{1/2}  \bm \Lambda_{\alpha}(\bm q,t) z_\alpha ||^2\rangle_{\bm q} > 0$ for an arbitrary non-zero vector~$\bm z $, positive definite matrix~$\mathsf B$, and non-zero vector~$\bm \Lambda_\alpha$. Then, the optimal condition for $\Delta {\hat S}^{(\ell)}(\bm k)$ can be written as
\begin{widetext}
\begin{align}
    \nabla_{\bm k } \Delta {\hat S}^{(\ell)}(\bm k)
    = \frac{ 2(\bm k^{\textsf T}  \langle \bm \Theta^{(\ell)} (\tau) \rangle  )[ (\bm k^{\textsf T}  \mathsf L^{(\ell)}(\tau ) \bm k)  \langle \bm \Theta^{(\ell)}(\tau ) \rangle - (\bm k^{\textsf T} \langle \bm \Theta^{(\ell)} (\tau)\rangle )(  \mathsf L^{(\ell)}(\tau ) \bm k ) ] }{ [\bm k^{\textsf T}  \mathsf L^{(\ell)}(\tau ) \bm k ]^2 } =0 ,  \label{eqA:opt_condition}
\end{align}
\end{widetext}
which is similar to Eq.~\eqref{eq:optimal}. The solution of Eq.~\eqref{eqA:opt_condition} is $\bm k^* =  (\mathsf L^{(\ell)}(\tau))^{-1} \cdot \langle \bm \Theta^{(\ell)}(\tau )\rangle$. By plugging this $\bm k^*$ into Eq.~\eqref{eqA:MEB_combined1}, we obtain the component-combined MEB as
\begin{align}
    \Delta S^{\rm tot} \geq&  \Delta\hat S^{(\ell)}(\bm k^*) = \langle \bm \Theta^{(\ell)} (\tau)\rangle   (\mathsf L^{(\ell)}(\tau))^{-1}   \langle \bm\Theta^{(\ell)}(\tau)\rangle .
\end{align}
This is the integral form of the component-combined MEB.
Following a similar way, we can derive the rate form of the component-combined MEB as
\begin{align}
    \sigma^{\rm tot}(t) \geq \langle \dot{\bm \Theta}^{(\ell)} (\tau)\rangle^{\textsf T}   (\dot{\mathsf L}^{(\ell)}(\tau))^{-1}  \langle \dot{\bm\Theta}^{(\ell)}(\tau)\rangle ,
\end{align}
where the components of $\langle \dot{\bm \Theta}^{(\ell)} (\tau)\rangle$ and $\dot{\mathsf L}^{(\ell)}(\tau)$ are given by
\begin{align}
    \langle \dot \Theta^{(\ell)}_\alpha (t) \rangle =&\int d \bm q \bm \Lambda_{\alpha} (\bm q,t )^{\textsf T}  \bm J^{\rm irr} (\bm q,t ),   \\
     (\dot{\mathsf L}^{(\ell)} (t ))_{\alpha,\beta} =& \langle \bm \Lambda_\alpha(\bm q,t )^{\textsf T}  \mathsf B  \bm \Lambda_\beta(\bm q,t )  \rangle_{\bm q}.
     \label{eq:ncwL}
\end{align}
Similar to Eq.~\eqref{eq:obtainL}, in the case where the estimation of the diffusion matrix requires a heavy computational cost, we can directly obtain $\dot{\mathsf L}^{(\ell)} (t )$ by evaluating
\begin{align}
    (\dot{\mathsf L}^{(\ell)} (t ))_{\alpha,\beta}=& \langle \bm \Lambda_\alpha (\bm q, t )^{\textsf T}  \mathsf B  \bm \Lambda_\beta (\bm q, t) \rangle \nonumber \\
    =& \lim_{\delta t \rightarrow 0} \frac{1}{\delta t} \langle[ \bm \Lambda_\alpha(\bm q,t)^{\textsf T} \circ \delta\bm q)][
    \bm \Lambda_\beta(\bm q,t)^{\textsf T} \circ \delta \bm q] \rangle .
\label{eq:NCWLalter}
\end{align}
Then, $({\mathsf L}^{(\ell)} (\tau ))_{\alpha,\beta}$  can be estimated by integrating Eq.~\eqref{eq:NCWLalter} over time from $t=0$ to $t=\tau$.

\section{Effect of limited samples or time resolution on the EP estimation }
\label{appendix:RNAlimiteddata}

\subsection{Limited number of trajectories }
\begin{figure}
\centering
\includegraphics[width=0.48\textwidth]{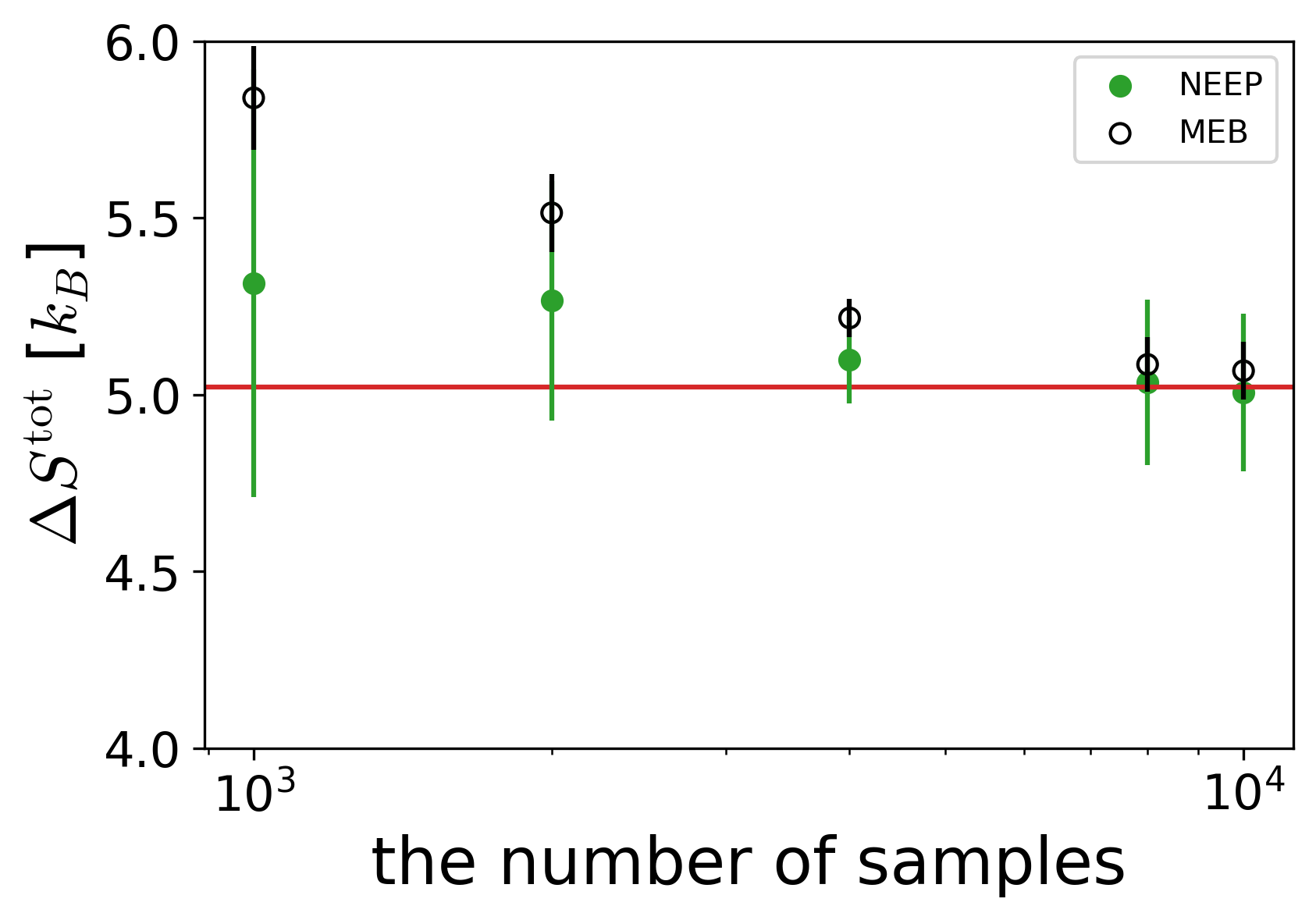}
\vskip -0.1in
\caption{Estimated EP of the RNA unfolding process as a function of the number of trajectory samples. Open circles represent the MEB results with 4 Gaussian bases ($\ell =4$). Green filled circles denote the NEEP results. The red solid line is the estimated EP via the direct method, which is the same line as in Fig.~\ref{fig:RNA}(c). The other parameters are the same as those used to plot Fig.~\ref{fig:RNA}(c). The error bars represent the standard deviation of the EP estimations from five different trajectory sets. }
\label{fig:EPversusNsamples}
\vskip -0.1in
\end{figure}

Though we use $10^5$ trajectories in our simulation in Sec.~\ref{sec:RNA}, only several thousand repetitions are usually feasible in real experiments~\cite{Jun2014}. Thus, in this section we examine the effect of a limited number of trajectories on the estimated EP. To this end, we perform additional simulations of the RNA unfolding process for various trajectory numbers of 1000, 2000, 4000, 8000, and 10000 and estimate the EP using both the MEB and NEEP methods. The results are plotted in Fig.~\ref{fig:EPversusNsamples}. Open and filled circles denote the MEB and NEEP results, respectively. 
To plot the error bars, we first generated 5 independent data sets for each number of samples, and then evaluated the average and standard deviation for the 5 sets. The red solid line is the estimated EP via the direct method, which is the same line as in Fig.~\ref{fig:RNA}(c). 
The figure shows the tendency that both MEB and NEEP overestimate the EP for small sample sizes. In fact, we recommend using both methods together to cross-check the reliability of the estimated value.

\subsection{Limited time resolution}
\begin{figure}
\centering
\includegraphics[width=0.48\textwidth]{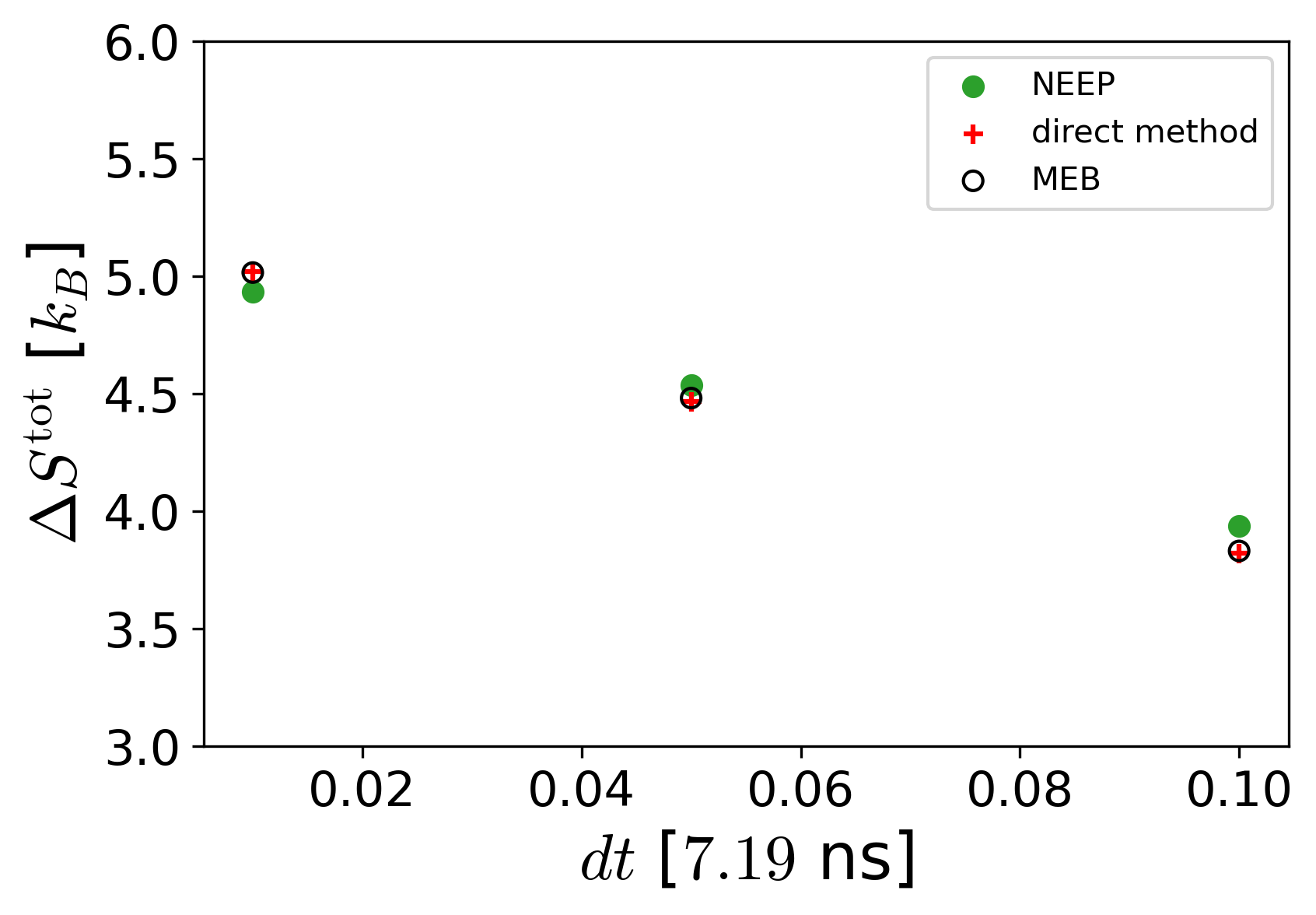}
\vskip -0.1in
\caption{Estimated EP as a function of time resolution~$dt$. Open circles represent the MEB results with 4 Gaussian bases ($\ell =4$). Green circles and red crosses denote the results of the NEEP and the direct method, respectively. The other parameters are the same as those used to plot Fig.~\ref{fig:RNA}(c). }
\vskip -0.1in
\label{fig:EPvsRes}
\end{figure}

The estimated EP depends on the time resolution $dt$ (time gap between two consecutive data points) of the measurement. Since decreasing the time resolution (increasing $dt$) causes a `coarse-graining' of the trajectory data, doing so typically leads to a lower value of the EP. This can also be checked in our simulations; here, we simulate the RNA unfolding process presented in Sec.~\ref{sec:RNA} with various $dt$. Other parameters are the same as those in the previous simulation. The results are plotted in Fig.~\ref{fig:EPvsRes}. The NEEP, the MEB, and the direct method show the same decreasing behavior as $dt$ increases. From an extrapolation of the data, we can also estimate the EP value in the $dt \rightarrow 0$ limit. Accordingly, it is important to specify the $dt$ information in a given experiment or simulation.

\section{NEEP algorithm}
\label{appendix:NEEP}

Here we explain the training details of the NEEP and its architecture configurations~\cite{kim2020neep}. We apply the NEEP to one step from $\bm x_{t}$ to $\bm x_{t+\Delta t}$. For brevity, we will use the notation $\bm x_{t}$ for $\bm x(t)$.
The NEEP is designed to maximize the following objective function:
\begin{align}
C(\theta) \equiv \left< \Delta{S_\theta(\bm x_{t+\Delta t}, \bm x_{t}, t)} - e^{-\Delta S_\theta(\bm x_{t+\Delta t}, \bm x_{t}, t)}\right>
\label{eq:NEEP's cost function}
\end{align}
where $\Delta S_\theta$ is an antisymmetric function with respect to the exchange of $\bm x_t$ and $\bm x_{t+\Delta t}$ as
\begin{align}
\Delta S_\theta(\bm x_{t+\Delta t}, \bm x_{t}, t) = h_{\theta}(\bm x_{t}, \bm x_{t+\Delta t}, t) - h_{\theta}(\bm x_{t+\Delta t}, \bm x_t, t). \label{eqA:DeltaS}
\end{align}
In Eq.~\eqref{eqA:DeltaS}, the function $h_\theta$ is the output of a multi-layer perceptron (MLP) and $\theta$ denotes the trainable parameters of the MLP. The MLP has a scalar output unit and three hidden layers of 512 units with the rectified linear unit (ReLU) activation function. It is shown that $\Delta S_\theta= \Delta S^{\rm tot}$ with the optimized $\theta^*$ in Ref.~\cite{kim2020neep}.

In order to employ the cross-validation method, we split the trajectory data into 20\% for the validation set and 80\% for the training set. We train the MLP $h_{\theta}$ to maximize Eq.~\eqref{eq:NEEP's cost function} by using the Adam optimizer~\cite{kingma2014adam} with learning rate $10^{-4}$, batch size 4096, and weight decay $5\times10^{-5}$. Before feeding the input $(\bm x_{t+\Delta t}, \bm x_{t}, t)$ to the MLP, we normalize each element of trajectory data $\bm x$ by using the following equation:
\begin{align}
x^{(i)}_{t} \leftarrow (x^{(i)}_{t} - \text{mean}[x^{(i)}]) / \text{std}[x^{(i)}],\nonumber
\end{align}
where $x^{(i)}$ indicates the $i$-th component of $\bm{x}$, $\text{mean}[x^{(i)}]$ is the mean of $x^{(i)}$, and $\text{std}[x^{(i)}]$ is the standard deviation of $x^{(i)}$. We also normalize the time information $t=0 \dots  \tau$ to be set as $t=-0.5 \dots 0.5$ so that mean of the input vector $(\bm x_{t+\Delta t}, \bm x_{t}, t)$ becomes a zero vector. The total number of training iterations is $10^4$, and we evaluate $C(\theta)$ values from the validation set per every 500 (50) training iterations for the pulled harmonic chain (RNA unfolding process). The best trained parameter set $\theta^\star$ is determined from the case where the NEEP produces the maximum value of $C$ during the training process. The results presented in Sec.~\ref{sec:verification} are those evaluated at the best trained parameter $\theta^\star$ over the total trajectory data.

\bibliography{reference}

\end{document}